\documentclass[useAMS,usenatbib]{mn2e}
\usepackage{graphicx}
\usepackage{float}
\usepackage{pstricks}
\usepackage{subfigure}
\usepackage{array}
\title{Two-shell collisions in the GRB afterglow phase}
\author[A. Vlasis, H. van Eerten, Z. Meliani, R. Keppens]{A. Vlasis$^{1,3}$\thanks{E-mail: 
Alkis.Vlasis@wis.kuleuven.be (AVR)}, H.J. van Eerten$^{5}$\thanks{E-mail:hve1@nyu.edu (AVR)}, Z. Meliani$^{1,3,6}$\thanks{E-mail: Zakaria.Meliani@wis.kuleuven.be}, R. Keppens$^{1,2,3,4}$\thanks{E-mail:
Rony.Keppens@wis.kuleuven.be (AVR)} \\ %R. Keppens$^{3}$\thanks{E-mail:email@address (AVR); otheremail@otheraddress (ANO)}
$^{1}$Center for Plasma Astrophysics, K.U.Leuven, Belgium\\
$^{2}$Astronomical Institute, Utrecht University, The Netherlands\\
$^{3}$Leuven Mathematical Modeling and Computational Science Centre, Leuven, Belgium\\
$^{4}$FOM institute for Plasma Physics Rijnhuizen, Nieuwegein, The Netherlands\\
$^{5}$Center for Cosmology and Particle Physics, Physics Department, New York University, New York, NY 10003\\
$^{6}$Observatoire de Paris, LUTH, F-92190 Meudon, France}

\begin{document} 

\date{Accepted ------. Received ------; in original form ------}

\pagerange{\pageref{firstpage}--\pageref{lastpage}} \pubyear{2009}

\maketitle

\label{firstpage}

\begin{abstract}
Strong optical and radio flares often appear in the afterglow phase of Gamma-Ray Bursts (GRBs). It has been proposed that colliding ultra-relativistic shells can produce these flares. Such consecutive shells can be formed due to the variability in the central source of a GRB. We perform high resolution 1D numerical simulations of late collisions between two ultra-relativistic shells in order to explore these events. We examine the case where a cold uniform shell collides with a self-similar Blandford and McKee shell in a constant density environment and consider cases with different Lorentz factor and energy for the uniform shell. We produce the corresponding on-axis light curves and emission images for the afterglow phase and examine the occurrence of optical and radio flares assuming a spherical explosion and a hard-edged jet scenario. For our simulations we use the Adaptive Mesh Refinement version of the Versatile Advection Code (AMRVAC) coupled to a linear radiative transfer code to calculate synchrotron emission. We find steeply rising flare like behavior for small jet opening angles and more gradual rebrightenings for large opening angles. Synchrotron self-absorption is found to strongly influence the onset and shape of the radio flare.
\end{abstract}


\section{Introduction}

\par The internal shock collisions version of the fireball model provides an adequate description of the origin of GRBs \citep{RM94}. A collapsing massive star \citep{Woos1993, WB2006} or a binary merger \citep{Narayan92} followed by a strong relativistic explosion are considered to be the progenitors of these violent events. Internal collisions inside the fireball of relativistic shells departing from the progenitor with different velocities, give rise to the GRB \citep{Piran04}. 

The same model attributes the afterglow emission to synchrotron radiation which is emitted during the deceleration of the external shock in the interstellar medium (ISM) \citep{Sari98, ZM04}. This behaviour can last for several days or even months after the burst covering a wide range of the spectrum. As afterglow observations improved, however, certain questions were raised that could not be answered with the standard model (\cite{Zhang07}, for a detailed discussion). Recent observations in the optical \citep{Stanek2006}, radio  as well as the X-ray band \citep{Burrows2005, Nousek2006}, show a strong variability in the afterglow phase for a large proportion of the bursts, which can not be reproduced by the standard external shock model.

It has been proposed that a bump in the afterglow light curve may result when the forward shock propagating in the ISM encounters a density jump, caused by an inhomogeneity of the surrounding medium generated by interstellar turbulence or by anisotropy in a precursor wind from the GRB progenitor \citep{Wang2000,Lazzati2002}. However, numerical simulations of a spherical explosion exhibit a rather canonical behaviour and even for a sharp and large increase in the external density this model does not produce sharp features in the light curve and cannot account for significant temporal variability in GRB afterglows \citep{NG2006,HvE09}. It has been suggested that a late activity of the central engine could explain the observed variability \citep{Falcone2006, Zang06, Romano2006, Ioka2005}.

In the late activity scenario, the central engine produces consecutive explosions after the initial burst which collide when a slow shell is followed by a faster one. This late activity of the source could be explained from a two-stage collapse in the central object. As proposed by \citet{King05} a collapsing core which has enough angular momentum to fragment will leave behind a second compact ``star'' in the form of a self-gravitating neutron lump. The fallback of this ``star'' at later times on the initial compact object can restart the central engine. Other theories suggest that the viscous hyperaccreting accretion disk around a black hole which fragments at large radii becomes dynamically unstable on different timescales and thus collapses at different times \citep{Perna2006}. It is proposed that the region at the vicinity of the accretor can play an important role in determining the accretion rate and therefore the energy output of the explosion \citep{Proga2006}.

According to late activity models the second blast wave continuously supplies the system with energy while colliding with the initially ejected material, producing in that way the rebrightening observed in the afterglow.

The role of magnetic fields in GRBs is still arguable. In the fireball model the presence of a magnetic field is not dynamically important for the evolution of the flow but plays an important role for the emission during the interaction of the flow with the external medium. Although the early afterglow emission strongly depends on the magnetization of the flow, in the late stages of the afterglow where the shells experience strong deceleration, the evolution of strongly magnetized shells resembles that of hydrodynamic shells and can be described by the self-similar Blandford-Mckee (BM) \citep{BM} approximation \citep{Mimica09}. At this stage of the afterglow the emission no longer contains information about the initial magnetization of the flow.

In section 2 we describe the high resolution numerical simulations we performed of late collisions of two ultra-relativistic shells during the afterglow phase. 
We claim that differences in the flow must have an impact on the resulting light curves and perform four simulations with varying Lorentz factor and energy content of the second shell in order to investigate the effect of these parameters. The adaptive mesh refinement (AMR) technique enables us to use high resolution in long term, 1D relativistic hydro simulations, in order to capture the forward and reverse shock formation on the second shell and study in detail the stages before, during and after the merger of the two shells. 

The effects of the collision between the two shells in the light curves is described in section 3. We study both spherical explosions as well as a hard-edged jet scenario where no lateral spreading has occured (numerical simulations in two dimensions have shown that only very modest lateral spreading occurs while the jet is relativistic \citep{ZMf09,Granot2001,Meliani2010}. Optical and radio on-axis light curves are calculated for different opening angles and the strength of the occuring flare or rebrightening is found to depend on this opening angle.
We also note a clear difference in the shape between optical and radio light curves as well as a difference in the time of the appearance of the flare between the two frequencies. We explain this chromatic behaviour in terms of the synchrotron self-absorption mechanism and the different main contributing regions of the jet to the emission.
We construct emission images for the different stages of the merger and connect the dynamical characteristics of the flow at each stage of the collision to the features in the light curves. 
We will discuss and summarize our results in section 4.

For the dynamical simulations we are using the Adaptive Mesh Refinement version of the Versatile Advection Code (AMRVAC) \citep{Kep03,Mel07}, and for the light curves and emission images calculations the radiation code of van Eerten \& Wijers (2009).

\section{Modeling of the multi-shell dynamics}

When the initially ultra-relativistic shell ejected from the central source starts to decelerate in the interstellar medium, a forward shock is created separating the shocked ISM from the ambient ISM. As mass is swept up, the kinetic energy of the shell is transformed into kinetic and thermal energy of the shocked matter. At the same time a reverse shock is formed which crosses the shell leading to conversion of the shell's kinetic energy into thermal. The resulting shocked ISM matter ultimately follows the self-similar BM analytical solution \citep{Mel07}. The afterglow is nowadays widely recognized as synchrotron radiation emitted during this phase of the propagation of the shell \citep{MR97}. 
% while certain variability in the early afterglow light-curves, such as optical and/or X-ray flares, has been attributed to synchrotron and synchrotron self-inverse Compton radiation from the reverse shock \citep{Kob07}. 

In our model we consider that the central engine remains active even after the initial ejection of the first shell resulting in a delayed second explosion. The produced blast wave will now travel with a steady velocity into an empty medium, since most of the matter has been swept up, until it reaches the termination shock of the first shell. In this paper we reproduce the collision process of these two shells and claim that for small opening angles the heating of the matter that happens during this phase is responsible for the appearance of the flares observed in the light curves. 

\subsection{Special relativistic hydrodynamic equations}

We perform the dynamical simulations using the 1D special relativistic hydrodynamic equations in spherical coordinates and the code AMRVAC.
The equations describing the motion of a relativistic fluid are given by the five conservation laws

\begin{equation}
(\rho u^{\mu})_{;\mu}=0, \ \ \ \ (T^{\mu\nu})_{;\nu}=0
\end{equation}
where $\mu , \nu = 0,1,2,3$ are the indices running over the 4-dimensional spacetime, $\rho$ is the proper rest mass density of the fluid, $u^{\mu}$ is the four-velocity and $T^{\mu\nu}$ is the stress-energy tensor given by $T^{\mu\nu}=\rho h u^{\mu}u^{\nu} + pg^{\mu\nu}$. Here with $p$ we denote the fluid rest frame pressure, while $g^{\mu\nu}$ is the Minkowski metric tensor, as we will consider a flat spacetime at distances far from the central engine. The specific enthalpy $h$ of the fluid is given by $h=1+ \varepsilon +p/\rho$ where $\varepsilon$ is the specific internal energy. Rewriting the conservation equations in vector form we have in a familiar 3+1 split the conservation laws
\begin{equation}
\frac{\partial U}{\partial t} + \frac{\partial F^{i}(U)}{\partial x^{i}}=0, \ \ \ \mbox{with} \ \ i=1,2,3.
\end{equation}
The vector $U$ is defined by the conserved variables as
\begin{equation}
 U= [D=\rho \gamma, {\bf{S}}=\rho h \gamma^{2} {\bf{v}}, \tau = \rho h \gamma^{2} -p -D]^{T},
\end{equation}
and the fluxes are given by 
\begin{equation}
 F=[\rho\gamma{\bf{v}}, \rho h \gamma^{2} {\bf{v}}{\bf{v}} + p{\bf{I}}, \rho h \gamma^{2}{\bf{v}}-\rho \gamma {\bf{v}}]^{T},
\end{equation}
where ${\bf{v}}$ is the three-velocity and $\bf{I}$ is the $3\times3$ identity matrix. The system of equations is closed by using the equation of state
\begin{equation}
 p=(\Gamma_{p} -1)\rho \varepsilon.
\end{equation}
In our simulations we choose the polytropic index to be $\Gamma_{p}=4/3$. This is a good approximation since most of the shocks in the cases described below are mainly relativistic or near-relativistic. 
% For the equations used in our simulations we have adopted units where the speed of light $c$ equals to unity.

\subsection{Initial setting}

For the dynamics of the first shell we consider that the reverse shock has already crossed the shell which now decelerates in the external medium. For the purpose of this simulation we will use the Blandford \& McKee (BM) approximation to describe this phase. This is a self-similar solution of a relativistic blast wave expanding in a uniform or radially varying medium. We consider the case where the explosion is assumed to be spherically symmetric and adiabatic.

In the BM model the density of the circumburst medium scales as a power law with distance $\rho_{1}(r)\propto r^{-k}$. For all the simulations in the present paper we will consider that the density of the circumburst medium is constant $(k=0)$, with particle number density $n_{1}=1\,\rm{cm^{-3}}$ and cold, with the pressure given by $p_{1}=10^{-5}n_{1}m_{p}c^{2}$ chosen such that it does not dynamically affect the system. We set the Lorentz factor of the BM shock at $\Gamma_{0}=23$ at the start of the simulation placed in distance $R_{0}\simeq2.04\times10^{17}$ cm. Considering the decelerating radius of the BM shock $R_{dec}=(3E_{0}/(4\pi n_{0}m_{p}c^{2}\Gamma^{2}))^{1/3}$, after which the Lorentz factor of the shock starts decreasing with distance as a power law, the initial distance $R_{0}$ of the shock corresponds to a distance 3.7 times greater than the deceleration radius $R_{dec}$ of a jet with initial Lorentz factor 100 and 7.8 times greater than the deceleration radius of a jet with initial Lorentz factor 300. The energy content of the shell is $E=10^{52}\,\rm{erg}$. According to the BM model in the ultra-relativistic case the jump conditions at the BM shock are given by

\begin{equation}
 p_{2}=\frac{2}{3} \Gamma^{2}\rho_{1},
\end{equation}

\begin{equation}
 n_{2}=\frac{2\Gamma^{2}}{\gamma_{2}}n_{1},
\end{equation}

\begin{equation}
 \gamma_{2}^{2}=\frac{1}{2}\Gamma^{2}.
\end{equation}
where index 2 denotes the shocked medium and $\Gamma$ is the Lorentz factor of the shock. According to the BM model in the ultra-relativistic case, the radius of the shock at time $t$ is given up to order $O(\Gamma^{-2})$ by  

\begin{equation}
 R(t) = ct\left(1-\frac{1}{8\Gamma^{2}}\right).
\end{equation}
From the jump conditions and by choosing the similarity variable to be $\chi=\left[1+8\Gamma^{2}\right]\left(1-r/t\right)$, we obtain the properties of the shocked medium 

\begin{equation}
p_{2}\left(r,t\right)=\frac{2}{3}\rho_{1}\Gamma^{2}\left[\left(1+8\Gamma^{2}\right)\left(1-\frac{r}{t}\right)\right]^{-17/12},
\end{equation}

\begin{equation}
 \gamma_{2}\left(r,t\right)=\frac{1}{2}\Gamma^{2}\left[\left(1+8\Gamma^{2}\right)\left(1-\frac{r}{t}\right)\right]^{-1},
\end{equation}

\begin{equation}
 \rho_{2}\left(r,t\right)=\frac{2\rho_{1}
\Gamma^{2}}{\gamma_{2}}\left[\left(1+8\Gamma^{2}\right)\left(1-\frac{r}{t}\right)\right]^{-7/4}.
\end{equation}
The total energy is then given by $E=8\pi \rho_{1}\Gamma_{0}^{2} c^{5}t_{0}^{3}/17$. If the initial Lorentz factor of the simulation is fixed at $\Gamma_{0}$, the duration $t_{0}$ of the shock so far then follows from this equation. The initial pressure and density jumps between the BM shell and the ISM at the position of the shock are $p_{2}/p_{1}= 10^{7}$ and $\rho_{2}/\rho_{1}=10^{2}$.

The second shell is uniform, cold and ultra-relativistic and placed at distance $\Delta R = 10^{14}$ cm behind the BM shell. It is therefore assumed that the shell has moved freely up to this point. Considering a duration of the second ejection event of $\Delta t = 1000$ s, the initial thickness of the shell will be $\delta = c\Delta t = 3\times10^{13}$ cm. The energy of the second shell is given by 

\begin{equation}\label{E_sh}
E_{sh}=4\pi\Gamma_{sh}^{2}R_{in}^{2}\delta\rho_{sh}c^{2},
\end{equation}
where $R_{in}$ denotes the initial distance of the shell and $\rho_{sh}$ and $\Gamma_{sh}$ the initial density and Lorentz factor. The initial pressure is chosen as $p_{sh}=5\times10^{-2}\rho_{sh}$. In our initial conditions, we vary from case to case the given parameters $\Gamma_{sh}$ and $E_{sh}$ which then serve to specify the shell density $\rho_{sh}$. The energy provided here refers to the isotropic-equivalent energy. For all cases described in this paper we consider emission along the rotation axis of the system and that the two ejecta have the same opening angle. We also neglect the effects of lateral spreading in both dynamical and radiation calculations. As shown in \citet{ZMf09}, sideways expansion can be a very slow process and definitely negligible for times under consideration in this paper. 
%A detailed 2D study is required to investigate both the effects of different opening angles of the two ejecta, as well as the spreading that may occur when the opening angle of the jet is extremely small, as in the case of $\theta_{jet}=2$ degrees considered later in this paper. 

We perform four simulations with varying Lorentz factor and energy for the second shell. In this simulation we are using a domain of size $[0.01, 10]\times10^{18}$ cm and 240 cells at the coarsest level of refinement. The physical properties of the afterglow shock collision model require a large domain and a very thin second shell which demands very high resolution in order to be resolved. The maximum level of refinement is 22, leading to an effective resolution of $5.03316\times10^{8}$ cells. We force the front and the back of the uniform shell to be refined in order to avoid numerical diffusion which would cause an artificial spreading of the shell. A summary of the simulation parameters can be found in Table 1. The top left figures in Figs. \ref{c9sn} and \ref{c12sn} show the initial conditions of case 1 and 4 respectively.

\begin{table}
\centering
\caption{Properties of the second shell for each case. $\Gamma=23$ and $E=10^{52}$ ergs are the Lorentz factor and energy of the BM shell.}
\begin{tabular}{cccr}
 \hline
 \hline
 case 1 & case 2 & case 3 & case 4 \\
 \hline
 \hline
   $\Gamma/\sqrt{2}$ & $2\Gamma/\sqrt{2}$ & $\Gamma/\sqrt{2}$ & $2\Gamma/\sqrt{2}$ \\
   $E$ & $E$ & $2E$ & $2E$ \\
 \hline
\end{tabular}
\label{table1}
\end{table}

\subsection{Interaction between two shells.}

\subsubsection{Jump conditions.}

The evolution of the two-shell system is shown in figures \ref{c9sn} and \ref{c12sn} for case 1 and case 4 respectively.  In case 1 the two shells initially have the same energy and Lorentz factor. As the BM shell decelerates in the interstellar medium the second shell catches up. While the matter from the BM shell is swept up by the second shell, a forward shock is created separating the shocked matter from the unshocked. At the same time a reverse shock crosses the second shell and a contact discontinuity appears in between both shocks. At this stage the front of the second shell has split into four regions. In region 1 there is the BM matter. In region 2 there is the BM matter which has been heated by the forward shock. In region 3 there is the matter of the uniform shell that has been heated from the propagation of the reverse shock and in region 4 there is the unshocked matter of the second shell. This is also shown for all 4 cases in Fig. \ref{etrans} where a close-up of the second shell at early times is given.

The jump conditions at the second shell as given by \citet{BM} for an arbitrary strong shock ($p_{2}/n_{2} \gg p_{1}/n_{1}$) are

\begin{equation}
e_{2}=\overline{\gamma}_{2}\frac{n_{2}}{n_{1}}w_{1}, \;\;\;\;\;\;\;\;\;\  \frac{n_{2}}{n_{1}}=4\overline{\gamma}_{2}+3,
\end{equation}

\begin{equation}
\frac{e_{3}}{n_{3}m_{p}c^{2}}=\overline{\gamma}_{3}-1, \;\;\  \frac{n_{3}}{n_{4}}=4\overline{\gamma}_{3}+3,
\end{equation}
where $\overline{\gamma}_{2}$ is the Lorentz factor of the fluid in region 2 relative to region 1 and $\overline{\gamma}_{3}$ is the Lorentz factor of the fluid in region 3 relative to region 4. The primitive variables $n_{i}$, $p_{i}$ as well as the internal energy density $e_{i}$ and enthalpy $w_{i}$, are measured in the fluid frame, while the Lorentz factors of the several regions, $\overline{\gamma}_{i}$, are measured in respect to the ISM which is considered to be at rest. In our case the forward shock of the second shell is moving in a hot medium which has already been heated by the BM shock. Therefore $w_{1}=e_{1}+p_{1}\simeq 4p_{1}$, assuming the ultra-relativistic equation of state $e_{1} \simeq 3p_{1}$. 

\subsubsection{Dynamics of the different cases.}

\begin{figure*}
 \centering
 \includegraphics[scale=0.4]{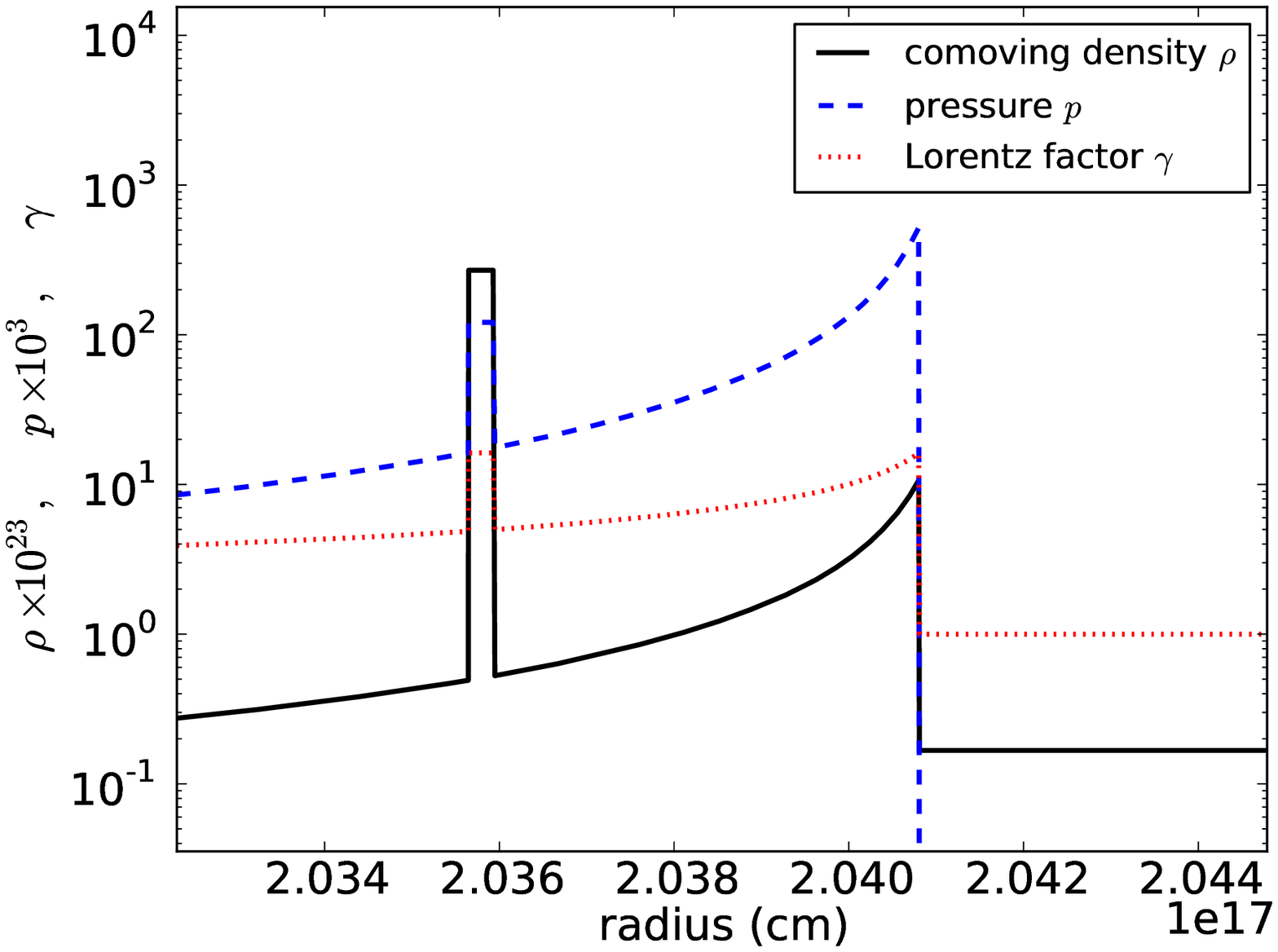}
 \includegraphics[scale=0.4]{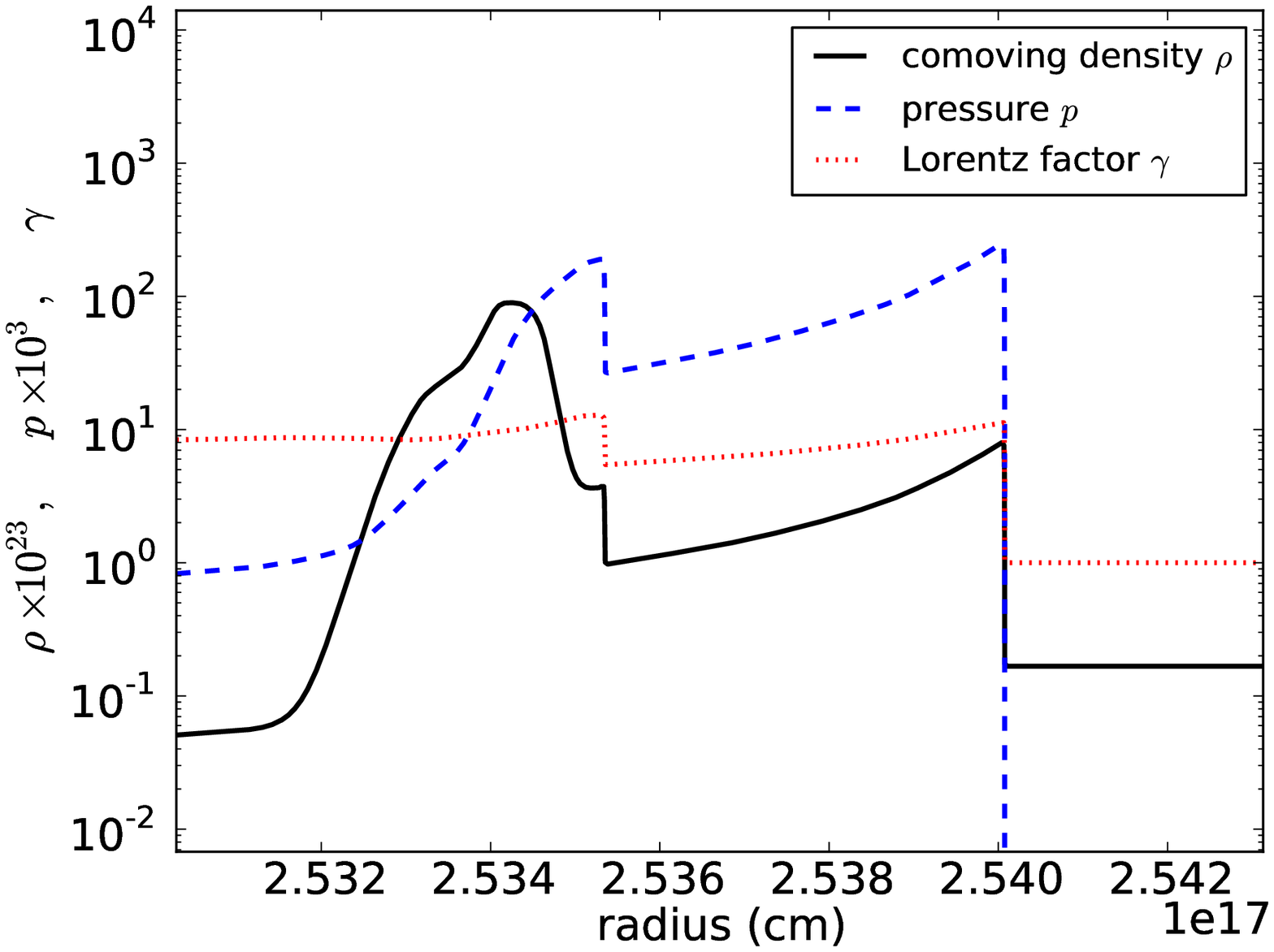}
 \includegraphics[scale=0.4]{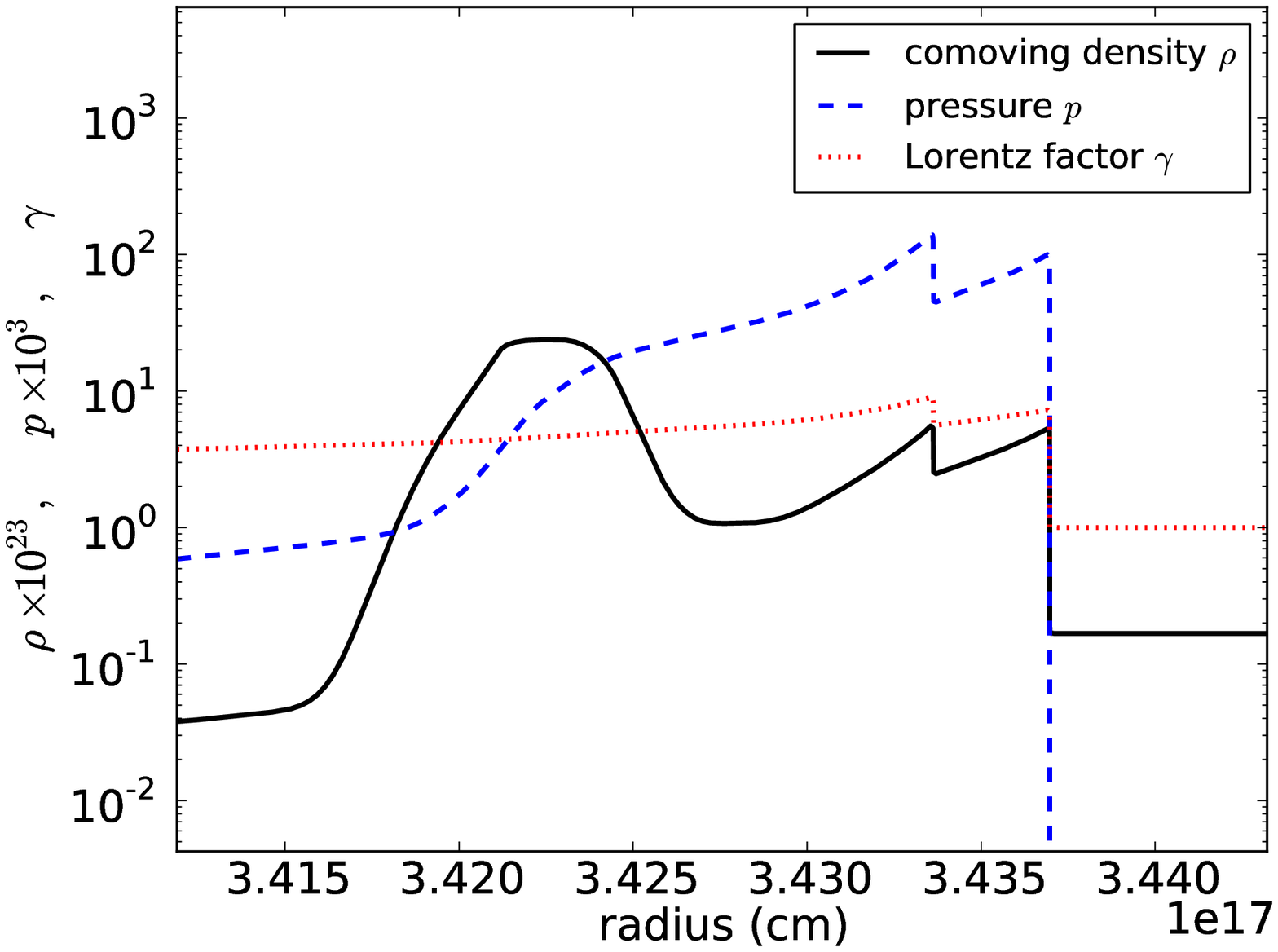}
 \includegraphics[scale=0.4]{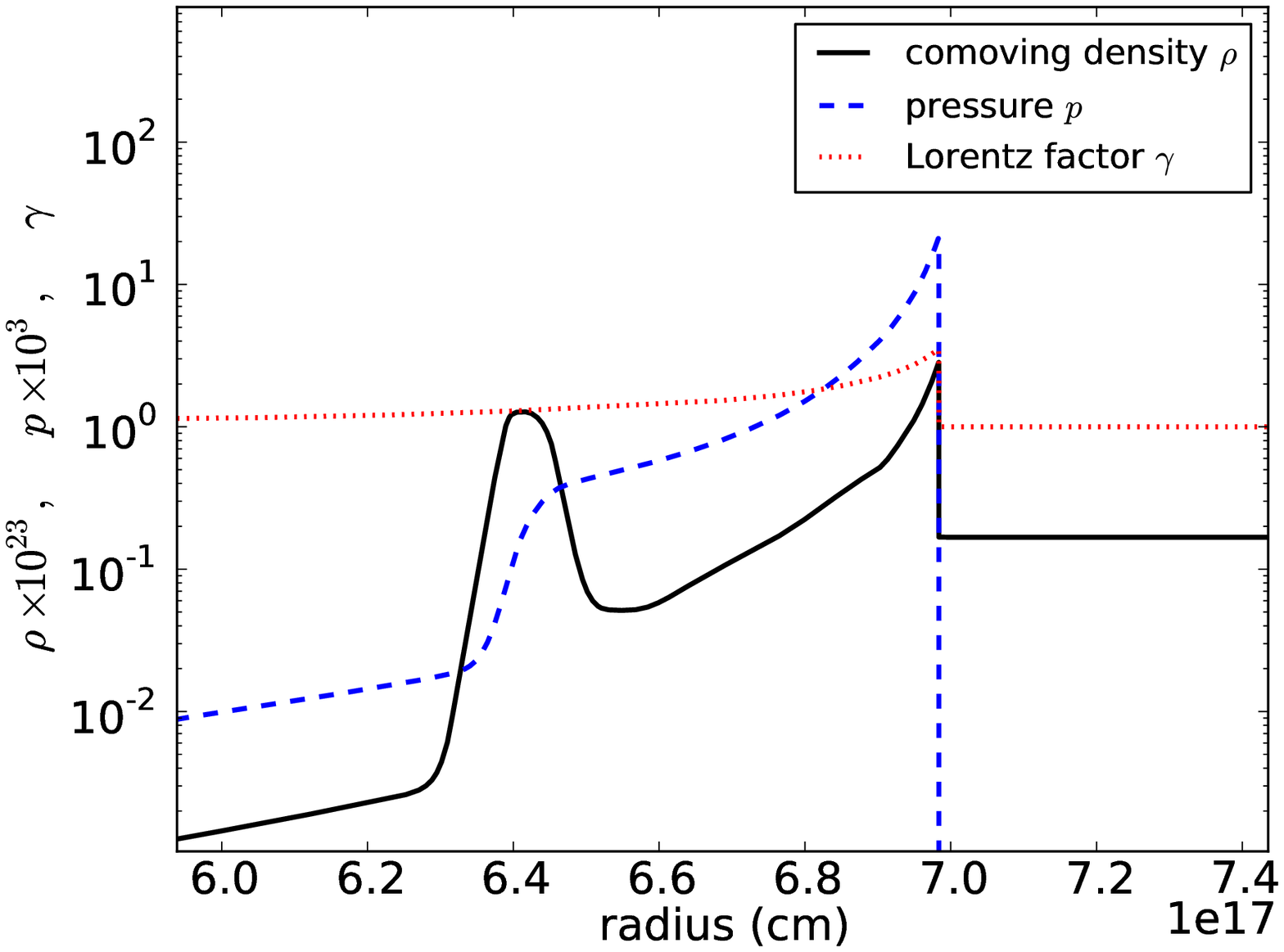}
\caption{Snapshots of the dynamics for case 1, taken at local emission times $t_{e}=6.81\times10^{6}$ s, $t_{e}=8.48\times10^{6}$ s, $t_{e}=1.15\times10^{7}$ s and $t_{e}=2.35\times10^{7}$ s after the initial explosion (top left - bottom right). Lorentz factor, density and pressure are indicated as the dotted, solid, and dashed line. The top-left figure indicates the initial setting of our simulation.}
\label{c9sn}
\end{figure*}

\begin{figure*}
 \centering
 \includegraphics[scale=0.4]{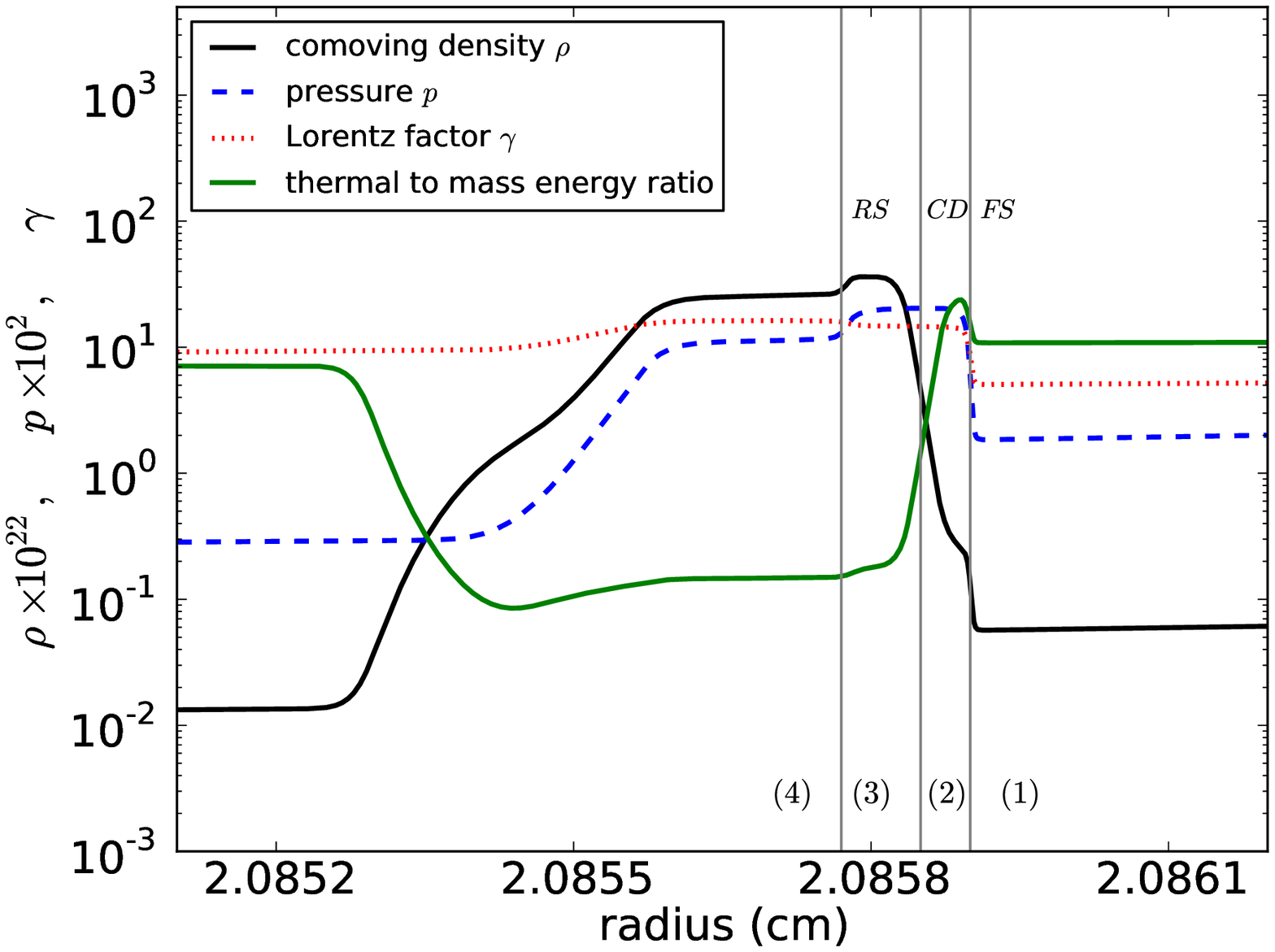}
 \includegraphics[scale=0.4]{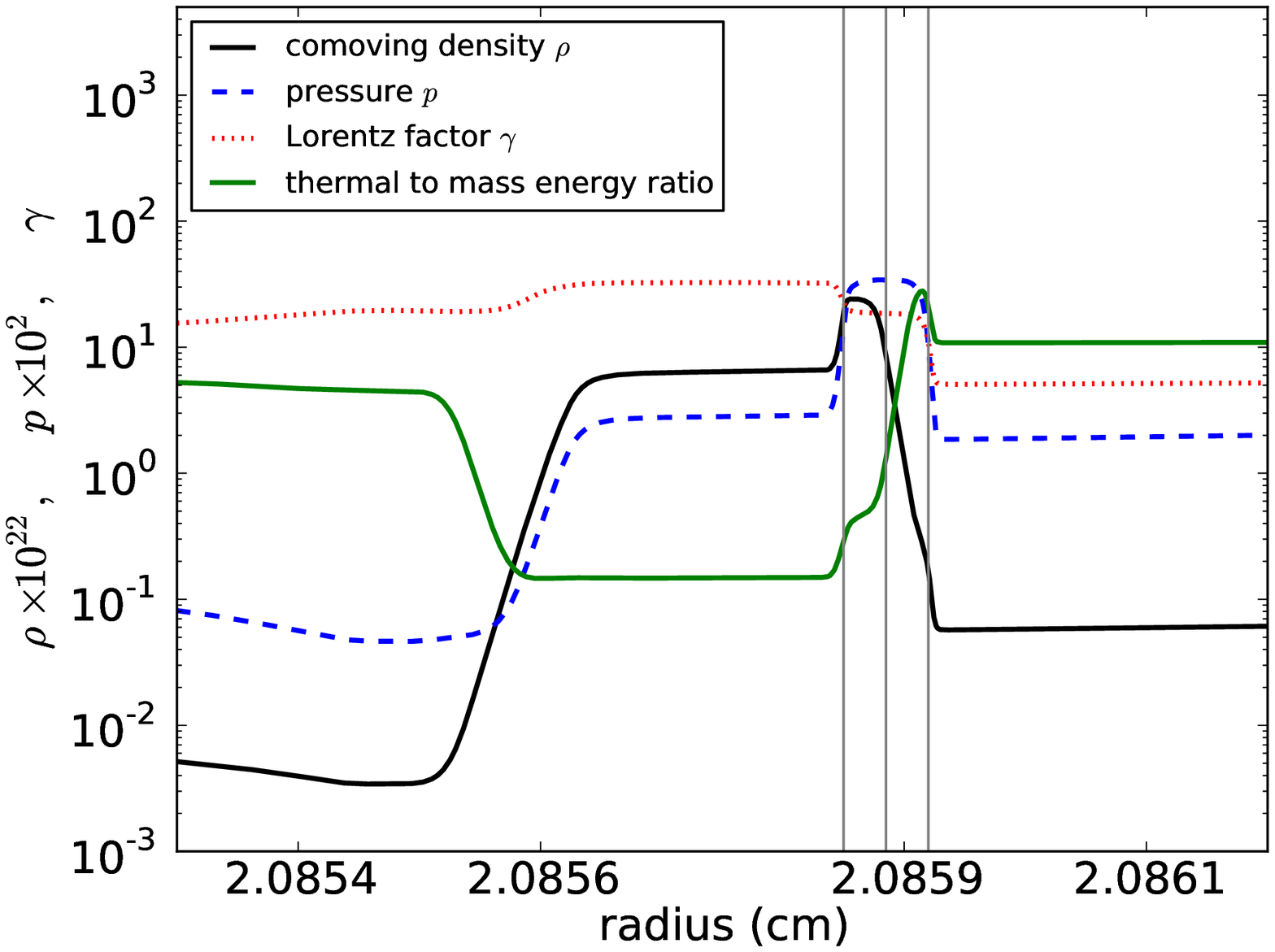}
 \includegraphics[scale=0.4]{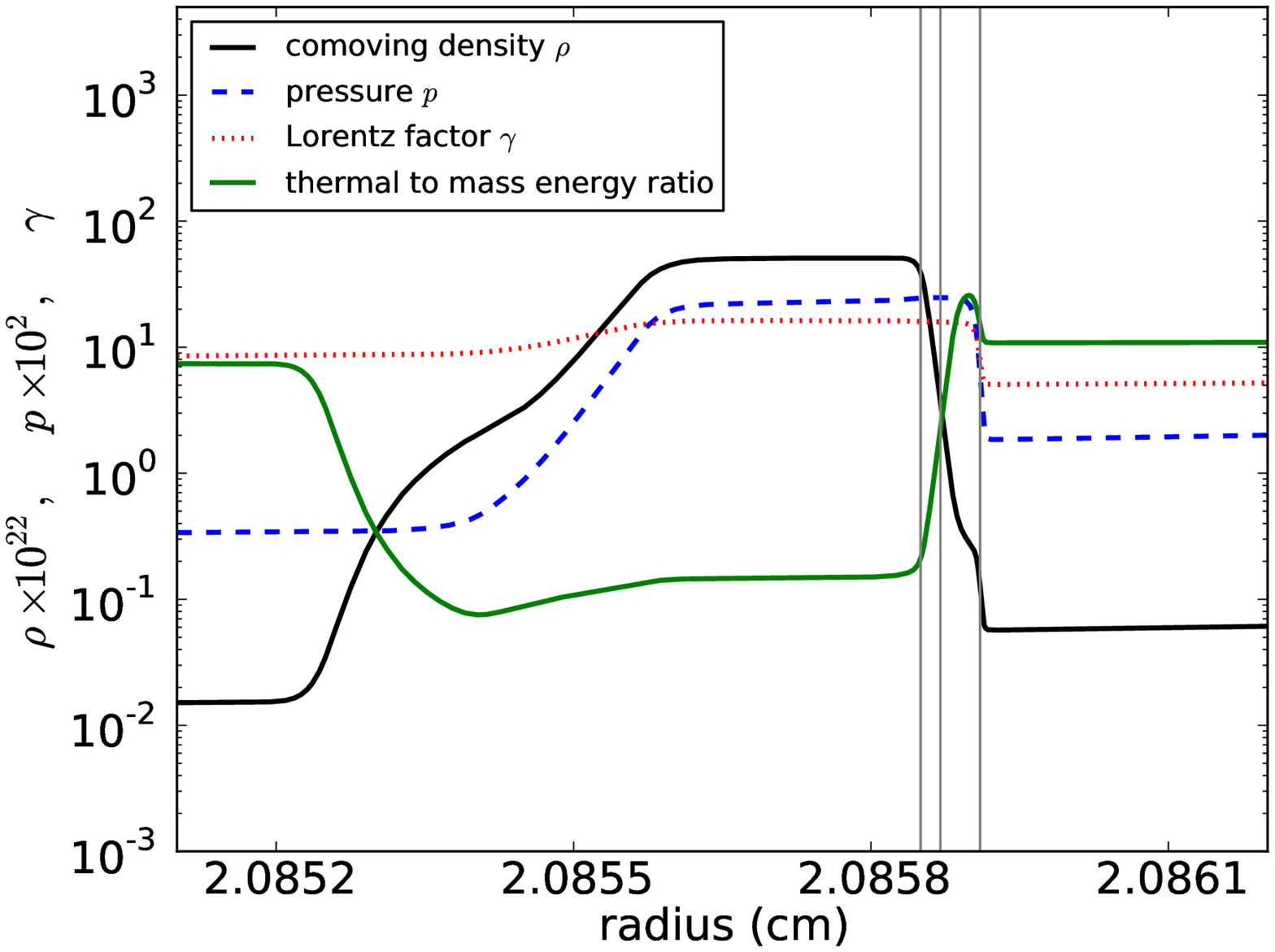}
 \includegraphics[scale=0.4]{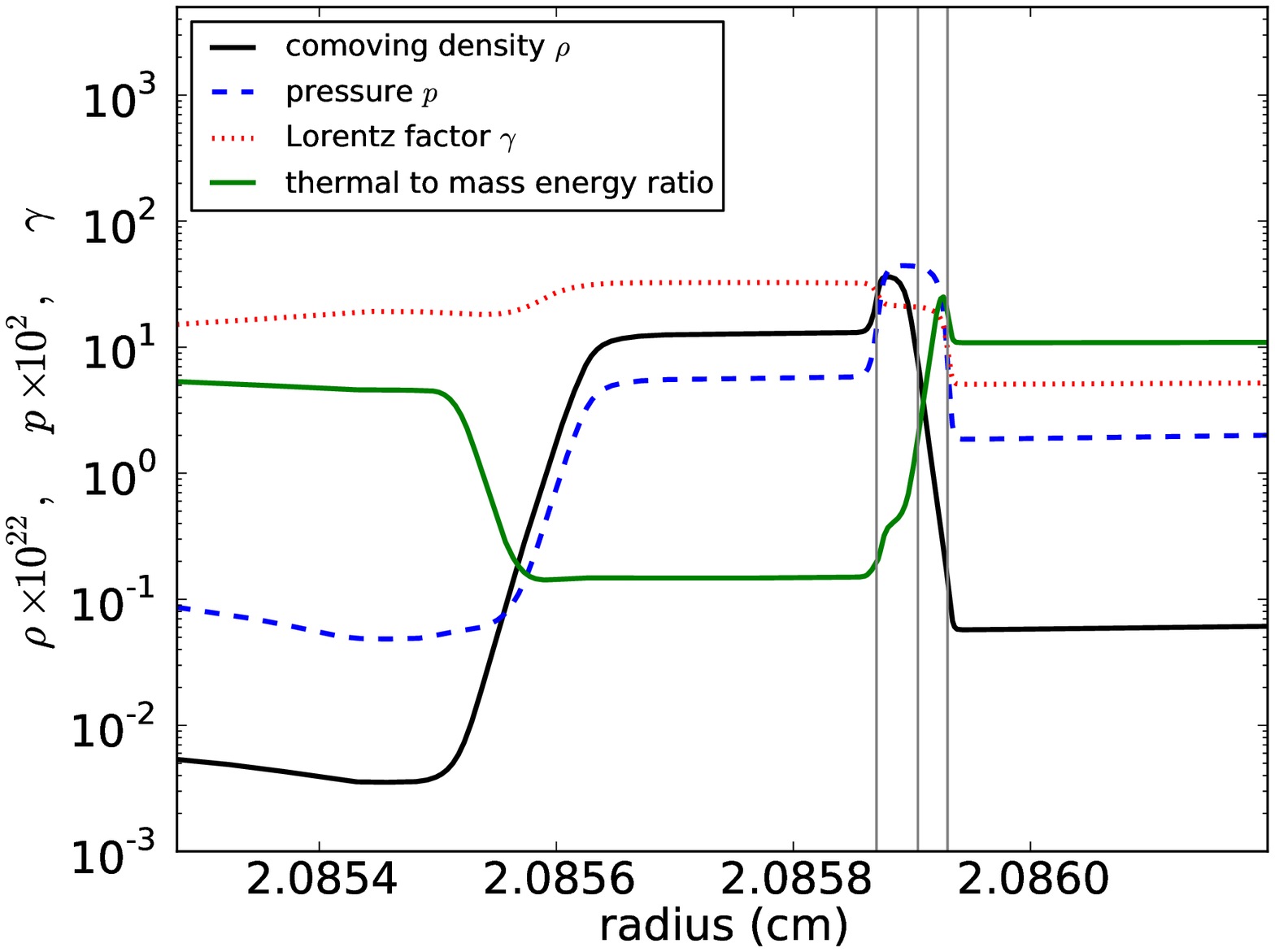}
\caption{Ratio of thermal to mass energy ($E_{th}/(\rho c^{2})$) and normalized Lorentz factor, density and pressure for case 1 to case 4 (top left to bottom right) for the second shell at emission time $t_{e}=6.98 \times 10^{6}$ s. Distance is normalized to $10^{17}$ cm. For cases 1 and 2 the transformation from thermal to kinetic energy is clear at the forward shock as well as the kinetic to thermal energy transformation at the position of the reverse shock. The ratio depends strongly on the properties of the second shell. The four regions of interaction between the two shells are indicated on the figure for case 1. Region (1) consists of the unshocked BM matter, region (2) of the shocked BM matter, region (3) of the shocked uniform shell matter and region (4) of the unshocked uniform shell matter. We focus on the front of the second shell, the dynamical evolution of which plays an important role on the light curves. }
\label{etrans}
\end{figure*}

After the initial explosion and acceleration of the uniform shell, the swept up BM matter starts playing an effective role in the deceleration of this shell. In Fig. \ref{etrans} we show snapshots of all 4 simulations taken at emission time $t_{e}= 6.98 \times 10^{6}$ s after the explosion when the forward shock, the contact discontinuity and the reverse shock are fully developed. At this time the BM matter is continuously heated by the forward shock resulting in an increase in the density of the shocked matter, $n_{2}/n_{1} \simeq 7.10$  (case 1). The relative Lorentz factor is given by 
$\overline{\gamma}_{2}=\gamma_{2}\gamma_{1}(1-\sqrt{(1-\gamma_{2}^{-2})(1-\gamma_{1}^{-2})}) \sim 1.46$. A small inconsistency is observed between the simulation results and the jump conditions which derives from the fact that in all our simulations the random kinetic energy per particle at the two sides of the shock is $p_{2}/n_{2}>p_{1}/n_{1}$ rather than $p_{2}/n_{2} \gg p_{1}/n_{1}$.
Similarly the propagation of the forward shock results in an increase of the internal energy of the shocked matter which now is 
$e_{2}\simeq 0.61 \sim \overline{\gamma}_{2}w_{1}n_{2}/n_{1}$.
At the same time the reverse shock crosses the shell while heating the uniform shell matter and transforming the kinetic energy of the shell into thermal. The contact discontinuity separating the two regions appears as a density jump between the two shocked fluids, $n_{3}/n_{2}\simeq 10^{2}$ while the Lorentz factor and pressure remain continuous, $\gamma_{2}=\gamma_{3}$ and $p_{2}=p_{3}$. The efficiency of this energy transformation mechanism is highly dependent on the reverse shock itself. 
According to Sari \& Piran (1995) the reverse shock depends only on two parameters, the Lorentz factor of the unshocked shell material relative to the BM unshocked matter, $\overline{\gamma}_{4}$, and the density jump, $n_{4}/n_{1}$. For $\overline{\gamma}_{4}^{2}\gg n_{4}/n_{1}$ the reverse shock will be relativistic, $\overline{\gamma}_{3} \gg 1$. For case 1, at emission time $t_{e}=6.98\times10^{6}$ s. (Fig. \ref{etrans}), the reverse shock remains Newtonian ($\overline{\gamma}_{3} = 1.00484$) while propagating into the shell and thus insufficient to heat the shell effectively. 
Following this behaviour, by the time the reverse shock reaches the back of the shell, the shell is still dominated by its kinetic energy although significantly decelerated (see also Fig. \ref{c9sn}).

We follow the same analysis for case 2 in which we double the Lorentz factor of the second shell compared to case 1. Since we choose to maintain constant the energy and the thickness of the second shell, the change in the Lorentz factor affects the density of the second shell, which is now smaller by a factor of 4 compared to case 1 (eq. \ref{E_sh}). Comparing the properties of the flow for the two cases at similar times after the explosion we notice that the forward shock is stronger in case 2. Specifically at emission time $t_{e}= 6.98 \times 10^{6}$ s after the explosion the forward shock has a Lorentz factor relative to the BM matter $\overline{\gamma}_{2} = 1.99$ and the number density satisfies $n_{2}/n_{1} \simeq 7.67$. That shows that the forward shock is more efficient in the second case as it compresses the matter of the BM shell to a higher degree and to higher Lorentz factor than in case 1. 
At the same time the reverse shock is stronger than in case 1, $\overline{\gamma}_{3} = 1.16$ and although it is not ultrarelativistic it is more efficient in converting the kinetic energy of the second shell into thermal. That appears clearly as a difference in the pressure between the shocked and unshocked shell matter, $p_{3}/p_{4}=12.27$ for case 2 compared to $p_{3}/p_{4}=1.78$ for case 1.

In case 3 we double the energy of the second shell. The relevant Lorentz factor between the shocked and unshocked BM matter is $\overline{\gamma}_{2} \simeq 1.75$ and the density jump satisfies $n_{2}/n_{1}\simeq 6.69$. Compared to case 1 the reverse shock for this case remains very weak, $\overline{\gamma}_{3}\simeq 1.00013$, while propagating in the uniform shell since it has to traverse a denser medium than before, leading to a compression ratio between the shocked and the unshocked shell matter $p_{3}/p_{4}=1.09$. In case 4 we double both the energy and Lorentz factor of the second shell. The initially very fast shell leads to a strong forward shock, $\overline{\gamma}_{2}\simeq 2.26$. The reverse shock is now stronger compared to case 3, $\overline{\gamma}_{3}\simeq 1.15$, compressing the shocked shell matter to higher pressure, $p_{3}/p_{4}=7.98$.

In figure \ref{etrans} we plot the thermal to mass energy ratio together with Lorentz factor, density and pressure for fixed early time for all cases. We observe that as matter from the BM shell is swept up from the forward shock of the second shell, the thermal energy of the fluid increases compared to the mass energy (which appears as a small bump in the thermal to mass energy curve at the position of the forward shock).
While the shell gets traversed by the reverse shock part of the kinetic energy of the shell is transformed into thermal as a result of the propagation of the reverse shock. We notice that the relativistic reverse shock in case 2 is a lot more efficient in raising the thermal energy component than in case 1. In case 3 and 4 a similar behaviour is observed. By increasing the energy in case 3 and with the Lorentz factor being the same as in case 1, the density becomes higher in the shell (eq. 13). As a result the reverse shock is very weak and highly inefficient in thermalizing the cold shell. In case 4 both the energy and the Lorentz factor of the second shell have twice the value compared to case 1. In this case the reverse shock is propagating in a less dense medium compared to case 3 and is now more efficient in transforming the kinetic energy of the second shell into thermal (Fig. \ref{etrans}). The Lorentz factor of the reverse shock is slightly overestimated as a result of our use of a fixed adiabatic index of 4/3 throughout the domain. Nevertheless, as we see from Fig. \ref{etrans}, the fluid is highly relativistic from the contact discontinuity of the second shell onward and therefore the regions that mostly contribute to the observed flux (see next section) are appropriately described by a fixed adiabatic index. 
%Although this holds for the shocked BM medium, an adiabatic index 5/3 for both the shocked and unshocked regions of the second shell would be more suitable.

\begin{figure*}
 \centering
 \includegraphics[scale=0.4]{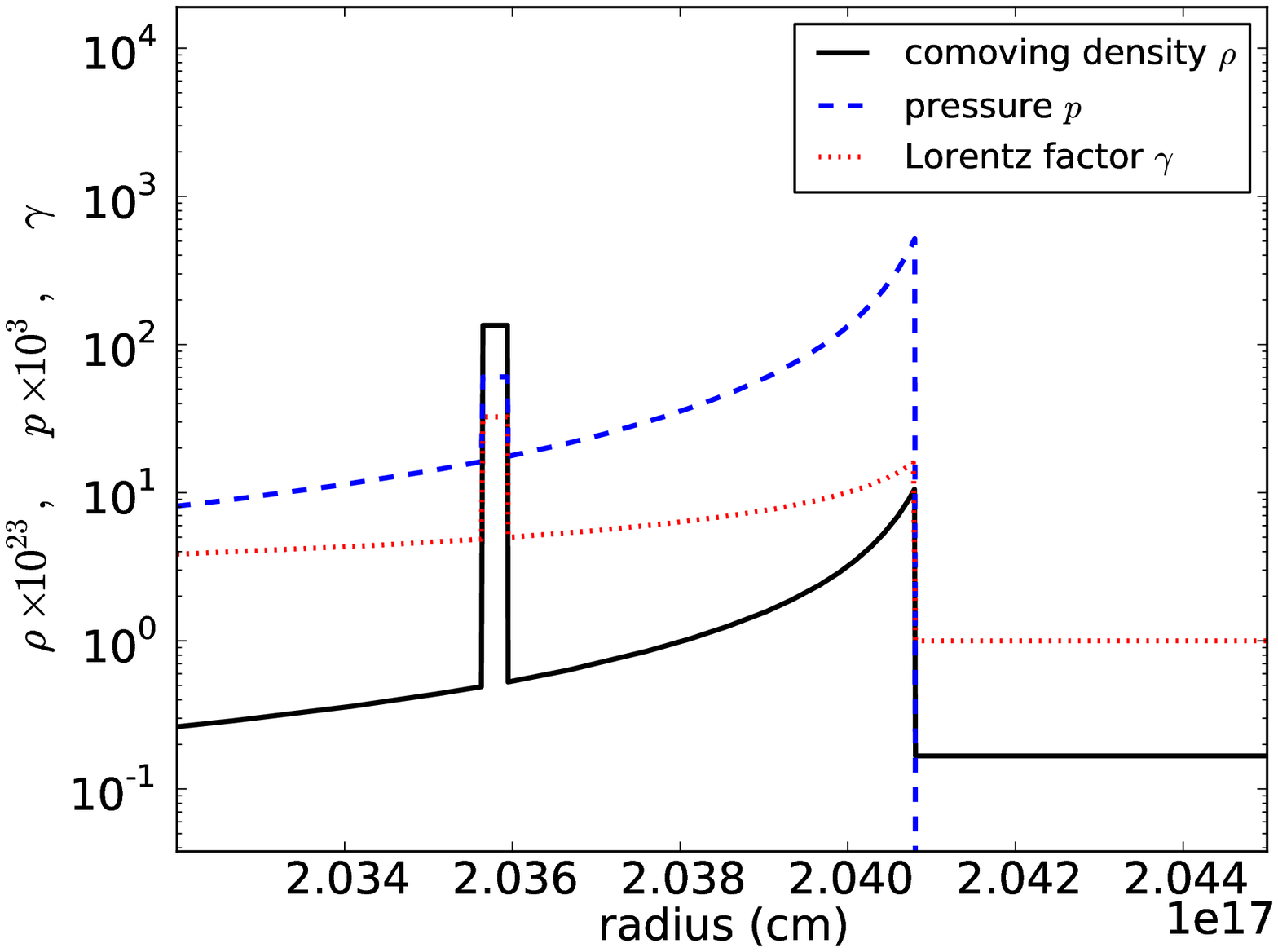}
 \includegraphics[scale=0.4]{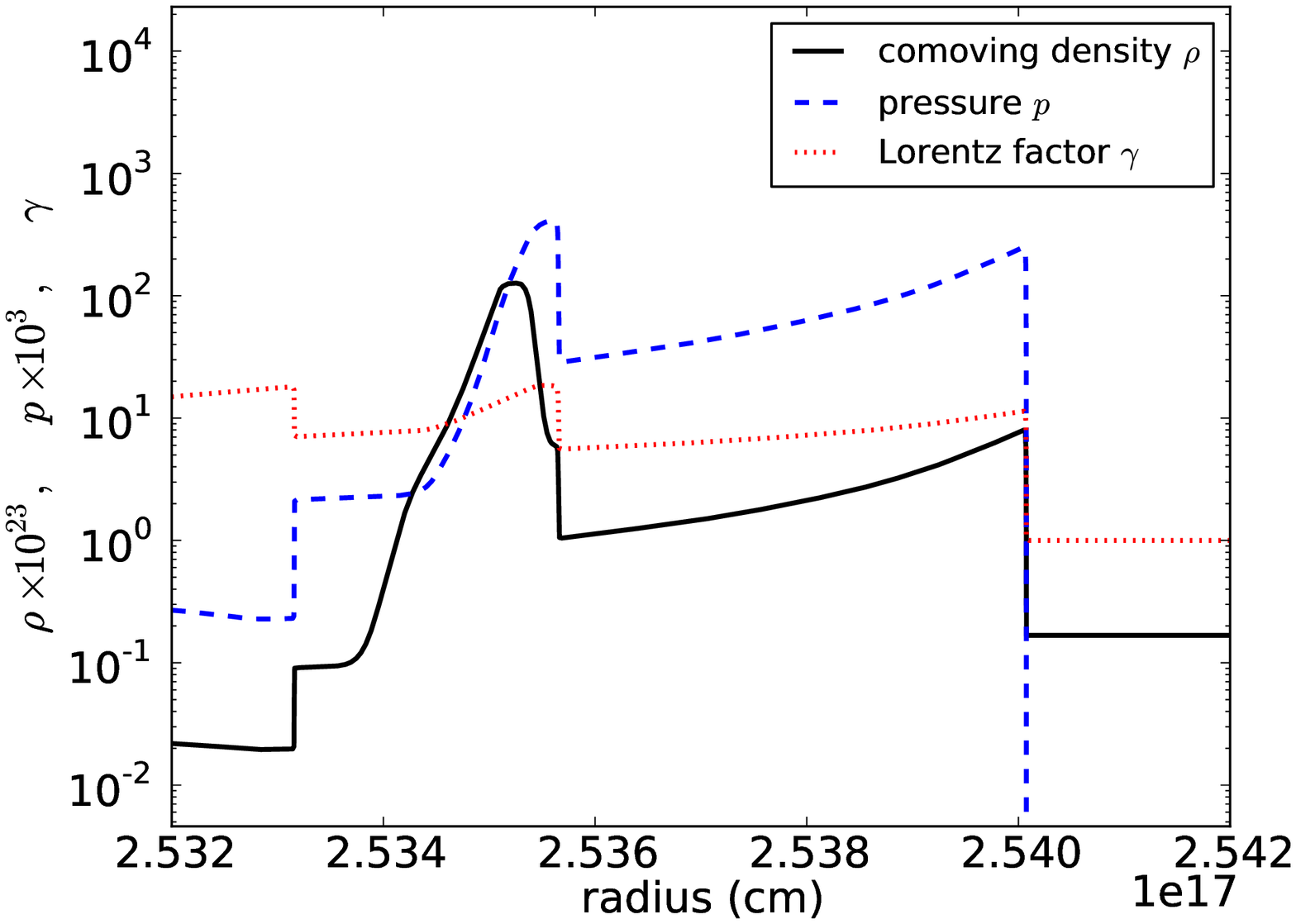}
 \includegraphics[scale=0.4]{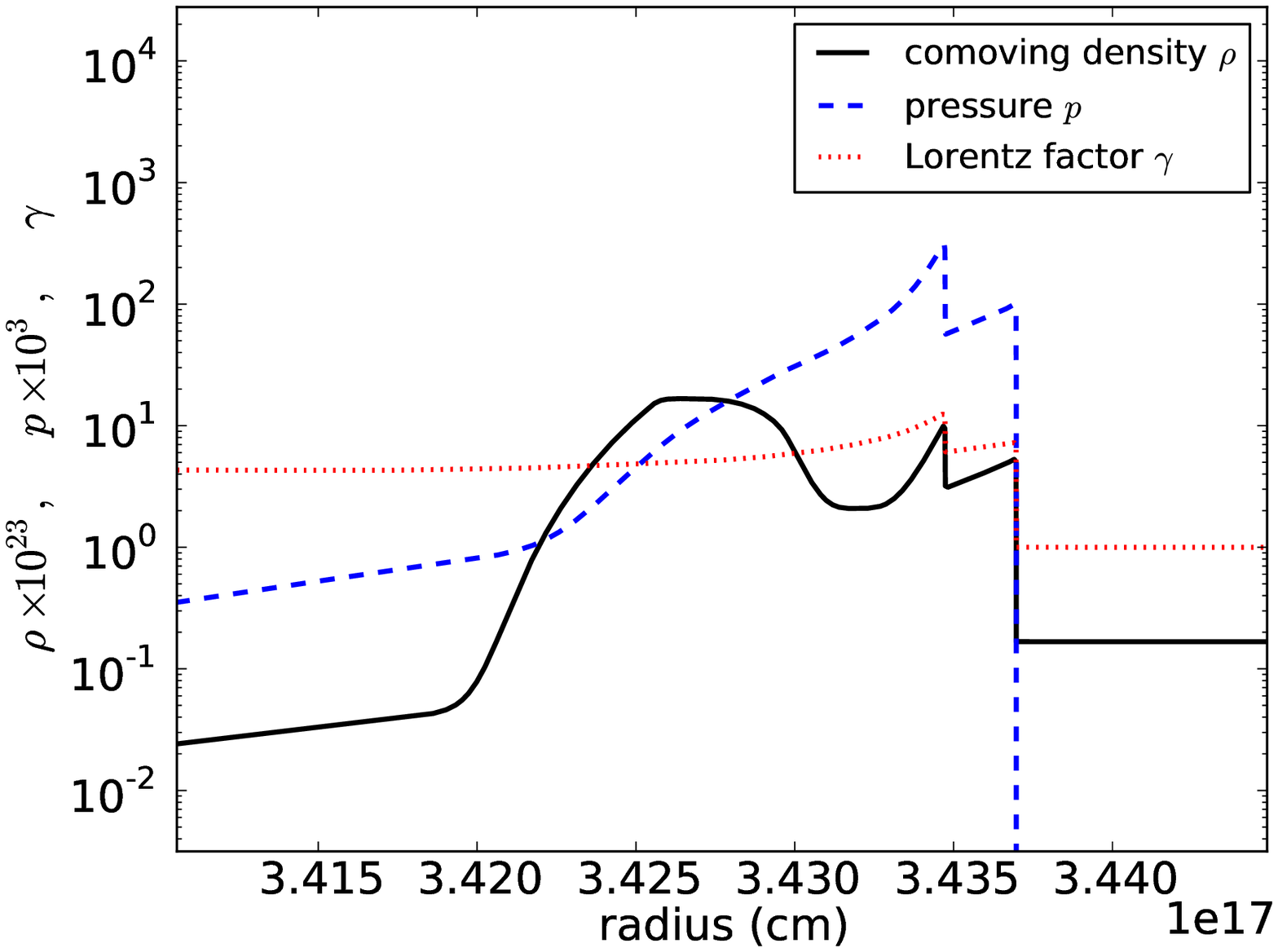}
 \includegraphics[scale=0.4]{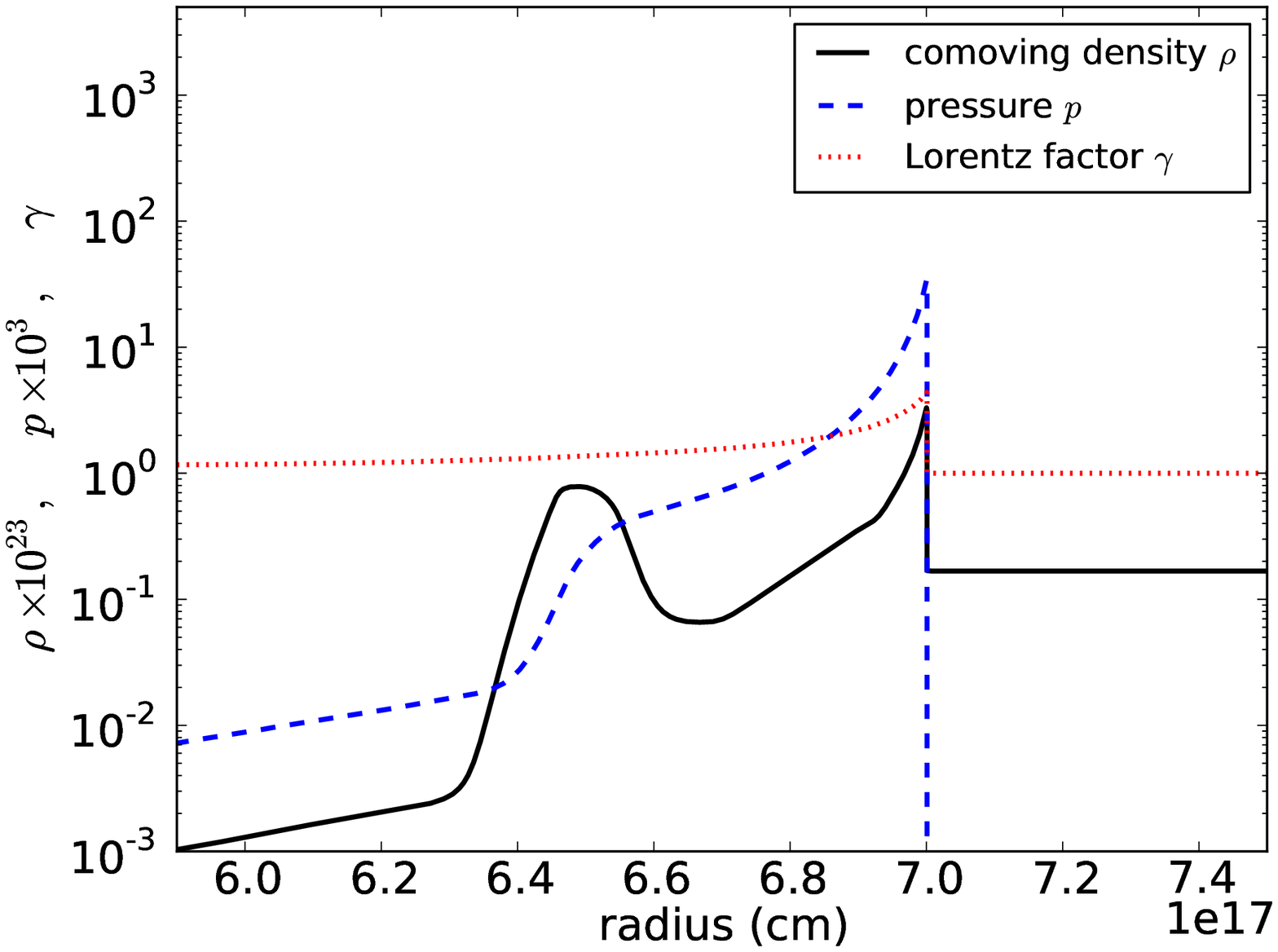}
\caption{Snapshots of the dynamics for case 4, taken at local emission times $t_{e}=6.81\times10^{6}$ s, $t_{e}=8.48\times10^{6}$ s, $t_{e}=1.15\times10^{7}$ s and $t_{e}=2.35\times10^{7}$ s after the initial explosion (top left - bottom right). Lorentz factor, density and pressure are indicated as the dotted, solid, and dashed line. The top-left figure indicates the initial setting of our simulation.}
\label{c12sn}
\end{figure*}

Later, at emission time $t_{e}=1.513\times10^{7}$ s (see Figs. \ref{c9sn} and \ref{c12sn}), the forward shock will catch up and overcome the BM shell as the latter one decelerates in the ISM. As the forward shocks merge and the reverse shock has crossed the back of the shell, a dense but slow and underpressured region is left behind unable to follow the forward shock as it has lost almost all of its kinetic energy reaching a near-equilibrium state with the surrounding matter.

\section{Radiation calculations}

In this section we describe the numerical calculations we performed in order to construct the light curves at the afterglow phase. The following calculations were carried out with the radiation code introduced in \citet{HvE09} and \citet{HvE10}. The main process contributing to the afterglow is synchrotron radiation. In both external and internal shock collision models, a magnetic field is required in order to fit the observational data. This magnetic field is most likely generated by instabilities forming in the shock such as the relativistic two-stream instability or the Weibel instability \citep{Med99, Weibel59}. In this case a fraction of the total thermal energy behind the shock goes to particle acceleration and another one to the generation of the magnetic field. Assuming that the fractions of the thermal energy density $e_{th}$, contributing to the magnetic energy density, $\epsilon_{B}$, and electron energy density, $\epsilon_{E}$, have a fixed value, we can calculate the afterglow emission. 
In our calculations we neglect the effect of the magnetic fields and radiative losses on the dynamics and do not take into account the effects of Compton scattering and electron cooling (that play a negligible role at times and frequencies under consideration). 
We do however include the effect of synchrotron self-absorption (ssa) due to the re-absorption of the radiation from the synchrotron electrons by solving simultaneously the linear radiative transfer equations for a large number of rays through the evolving fluid. An analytical formula for obtaining the self-absorption frequency $\nu_{sa}$ can be found in \citet{Granot1999b}.

The radiation code is specifically written to include the snapshots produced by the dynamical simulations performed with AMRVAC. In all our calculations the values of $\epsilon_{B}$ and $\epsilon_{E}$ are fixed to 0.01 and 0.1 respectively and we assume a power law distribution for the accelerated electrons with $p=2.5$. We also fix the fraction of the electrons accelerated to this power law distribution $\xi_{N}$ equal to 0.1. Assuming isotropic radiation in the comoving frame, the observer's flux calculated by the radiation code is given by 

\begin{equation}
 F_{\nu}= \frac{1+z}{d_{L}^{2}}\int \frac{d^{2}P_{V}}{d\nu d\Omega}(1 - \beta \mu)cdAdt_{e},
\end{equation}
in the optically thin limit, where for the purpose of our simulations the redshift z is chosen to be zero. Here $D_{L}$ denotes the observer luminosity distance, $\beta$ the fluid velocity in units of $c$, $d^{2}P_{V}/d\nu d\Omega$ is the received power per unit volume, frequency and solid angle and $\mu=\cos\theta$, with $\theta$ being the angle between the fluid velocity and the line of sight. The surface element $dA$ corresponds to an equidistant surface \textit{A}, which is a surface intersecting the fluid grid from which the radiation arrives at the observer at time $t_{obs}$. Each surface corresponds to a specific emission time $t_{e}$. For a photon emitted from a location (r,$\theta$) at emission time $t_{e}$, the observer time is given by

\begin{equation}
 t_{obs}= t_{e} - \frac{r\mu}{c}.
\end{equation}

When electron cooling does not play a role, the shape of the observed spectrum follows directly from the dimensionless function $Q(\nu / \nu_{m})$ which has the limiting behaviour $Q \propto (\nu/\nu_{m})^{1/3}$ for small $(\nu/ \nu_{m})$ and $Q\propto(\nu/\nu_{m})^{(1-p)/2}$ for large $(\nu/\nu_{m})$. The received power depends on this shape and on the local fluid quantities via

\begin{equation}
 \frac{d^{2}P_{V}}{d\nu d\Omega} \propto \frac{\xi_{N}nB'}{\gamma^{3}(1-\beta\mu)^{3}}Q\left(\frac{\nu}{\nu_{m}}\right),
\end{equation}
where $n$ is the lab frame number density of the electrons and $B'$ is the comoving magnetic field strength. The synchrotron peak frequency $\nu_{m}$ corresponds to the Lorentz factor of the lower cut-off of the accelerated electron's power-law distribution $\gamma_{m}$, assuming that the Lorentz factor for the upper cut-off goes to infinity. Then the lower cut-off for the electrons will relate to the comoving number density $n'$ and thermal energy density $e'_{th}$ via 

\begin{equation}\label{gammam}
\gamma_{m}\propto\left(\frac{p-2}{p-1}\right)\frac{\epsilon_{E}e'_{th}}{\xi_{N}n'}.
\end{equation}

The particle distribution and its lower cut-off are set at the shock front. However, subsequent evolution of $\gamma_{m}$ is dictated by adiabatic expansion of the fluid rather than synchrotron radiative losses. Therefore, in a relativistic fluid, eq. (\ref{gammam}) also holds further from the shock front with the same value of $\epsilon_{E}$.

The temporal behaviour of several GRBs contradicts the spherical explosion senario. A rapidly decaying afterglow emission suggests that the flow must be collimated rather than spherical. This is vital for the GRB mechanism since a spherical expansion would require a total energy budget $\sim 10^{54}$ ergs which is hard to produce from a stellar mass progenitor, while a jet shaped explosion can have the same result with less energy, $\sim 10^{51}-10^{52}$ ergs. This model suggests that when the jet decelerates to Lorentz factors such that $\gamma \sim \theta_{h}^{-1}$ is satisfied, with $\theta_{h}$ being the half jet opening angle, the flux that the observer receives will start decreasing resulting to a break in the afterglow light curves. In contrast to AGN jets that can be directly observed, GRB jets are only implicitly assumed from this break. 
 
In our simulations we construct optical and radio light curves and emission images at various times for the 4 cases described above. We try different opening angles of the jet and associate the flare characteristics with the jet opening angle. 

The main characteristic of our approach is the separation of the dynamics from the radiation simulation. The outputs from the dynamical simulations are used as an input to the radiation code in order to calculate synchrotron emission. When absorption plays a role, the code solves a series of linear radiative transfer equations (rather than eq. 16 directly) along the light rays starting from the back of the jet and passing through the jet towards the observer. For given emission time $t_{e}$ a surface on the radiative volume of the jet exists from which emission arrives at the observer at time $t_{obs}$.  
A summation over the light rays along this \textit{equidistant surface}, and an integration over all the \textit{equidistant surfaces} is applied in order to calculate the contribution to the emission of every part of the jet. The observed flux is obtained by summing over the rays emerging from the jet. When the rays are not summed over, a spatially resolved emission image of the two shell system is obtained. An adaptive procedure similar to the one used in AMRVAC is employed when more rays need to be calculated in order to adequately capture the emission from the underlying fluid profile.
More details on the radiation code algorithmic strategy can be found in \citet{HvE09, HvE10}.

\subsection{Early afterglow and jet break estimation.}

The dynamical simulations cover a timescale starting from 0.072 days and ending 10 days after the initial explosion in the observer's time frame. In order to include the initial stages of the afterglow in our simulation we assume that prior to the simulation the outer shock has evolved according to BM and the second has moved with a constant velocity while retaining its initial shape. 
In this way the deceleration of the initial explosion is taken into account resulting in the appearance of the jet break in the light curve. 

Due to relativistic beaming the emission that reaches the observer is limited to emission angles $\theta<\gamma^{-1}$. During the afterglow phase however the flow is significantly decelerated and thus larger emission angles can be observed. When the Lorentz factor becomes small enough so that the condition $\theta_{h}\sim\gamma^{-1}$ is satisfied, where $\theta_{h}$ is the half opening angle of the jet, the observation cone becomes big enough for the edges of the jet to become visible to the observer. When there is no significant spreading of the matter at the edges only part of the visible region is occupied by the jet and thus from this point after the flux will start decaying faster. This transition to the faster decaying part of the light curve is often referred to as the jet break. 

An analytical formula given in \citet{HvE10b} gives an estimation of the observed jet break time, $t_{obs,br}$ depending on the total energy of the explosion $E$, the half opening angle $\theta_{h}$ of the jet, the circumburst density $n_{0}$ and profile $k$. The contributing area to the emission in this case is not only the shock front itself but also the area behind the shock. For that purpose the radial profile from the BM model has been used, and the formula reads 

\begin{eqnarray}\label{tbreak}
\lefteqn{t_{obs,br}=\theta_{h}^{2+2/(3-k)}\left(\frac{A_{1}}{2(1+2(4-k))}\right)^{1/(3-k)}\times} \nonumber \\
  & & \left(\frac{1}{2}+\frac{1}{2(1+2(4-k))} + \frac{1}{2(4-k)2(1+2(4-k))}\right)
\end{eqnarray}
where $k$ corresponds to the parameter defining the power-law of the circumburst medium density, which for our case is always set to 0, and $A_{1}$ a parameter of the fluid depending on the energy $E$ of the explosion and the number density at the position of the shock, $A_{1}=E(17-4k)/(8\pi m_{p} n_{0} R_{0}^{k} c^{5-k})$ with $n_{0}$ being the proton number density at distance $R_{0}$.

\begin{table}
\centering
\caption{Observer time estimation of the jet break (estimation made in days after the explosion) for the four simulated cases assuming two hard-edged jet scenarios. }
\begin{tabular}{c|cccr}
 \hline
 \hline
\multicolumn{5}{c}{\bf{Jet break estimation}} \\
 \hline
 & case 1 & case 2 & case 3 & case 4 \\
 \hline
  $2\theta_{h}=2 $ & $0.0028$ & $0.0028$ & $0.0035$ & $0.0035$ \\
  $2\theta_{h}=5 $ & $0.032$ & $0.032$ & $0.041$ & $0.041$ \\
 \hline
 \hline
\end{tabular}
\label{table2}
\end{table}

The jet break observer time estimation $t_{obs,br}$, is presented in Table \ref{table2} for all the cases we simulate and for two different opening angles of the jet. As expected, when the total energy of the explosion remains the same, the jet break time remains the same even if the Lorentz factor of the second shell is different. For larger opening angles of the jet the jet break is observed at later observer times as it takes longer for the shell to decelerate to Lorentz factors such that $\theta_{h} \sim \gamma^{-1}$ condition is satisfied. We discuss the jet break characteristics further in section 3.4.

In our model the interaction between the two shells starts happening after the jet break has occured, leading to a rebrightening of the afterglow and a sudden increase in the flux. The shape of the light curve and the characteristics of the emission are subjected to the dynamics of the flow and present qualitative and quantitative differences from case to case.

\begin{figure*}
 \centering
  \includegraphics[scale=0.4]{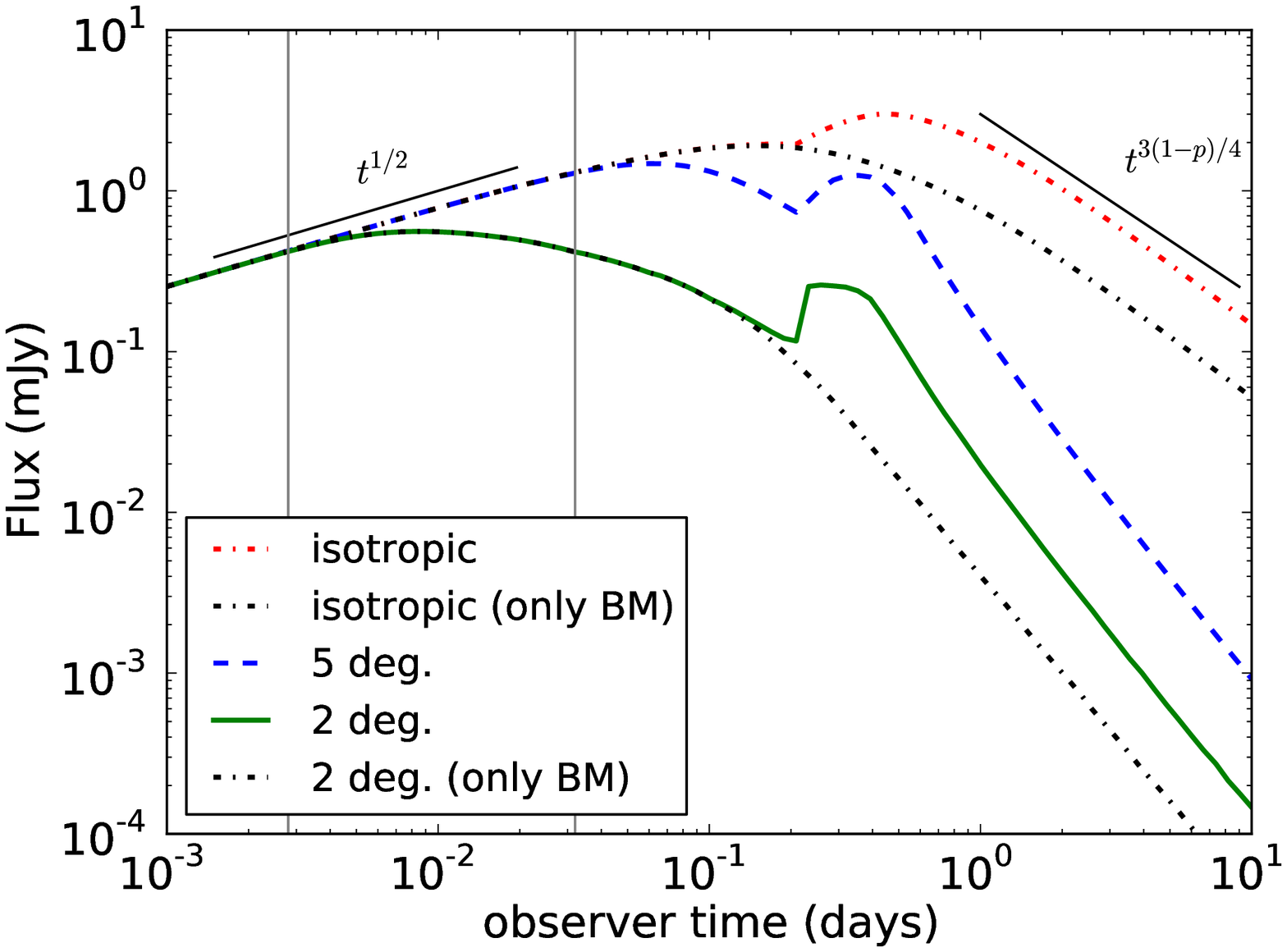}
 \includegraphics[scale=0.4]{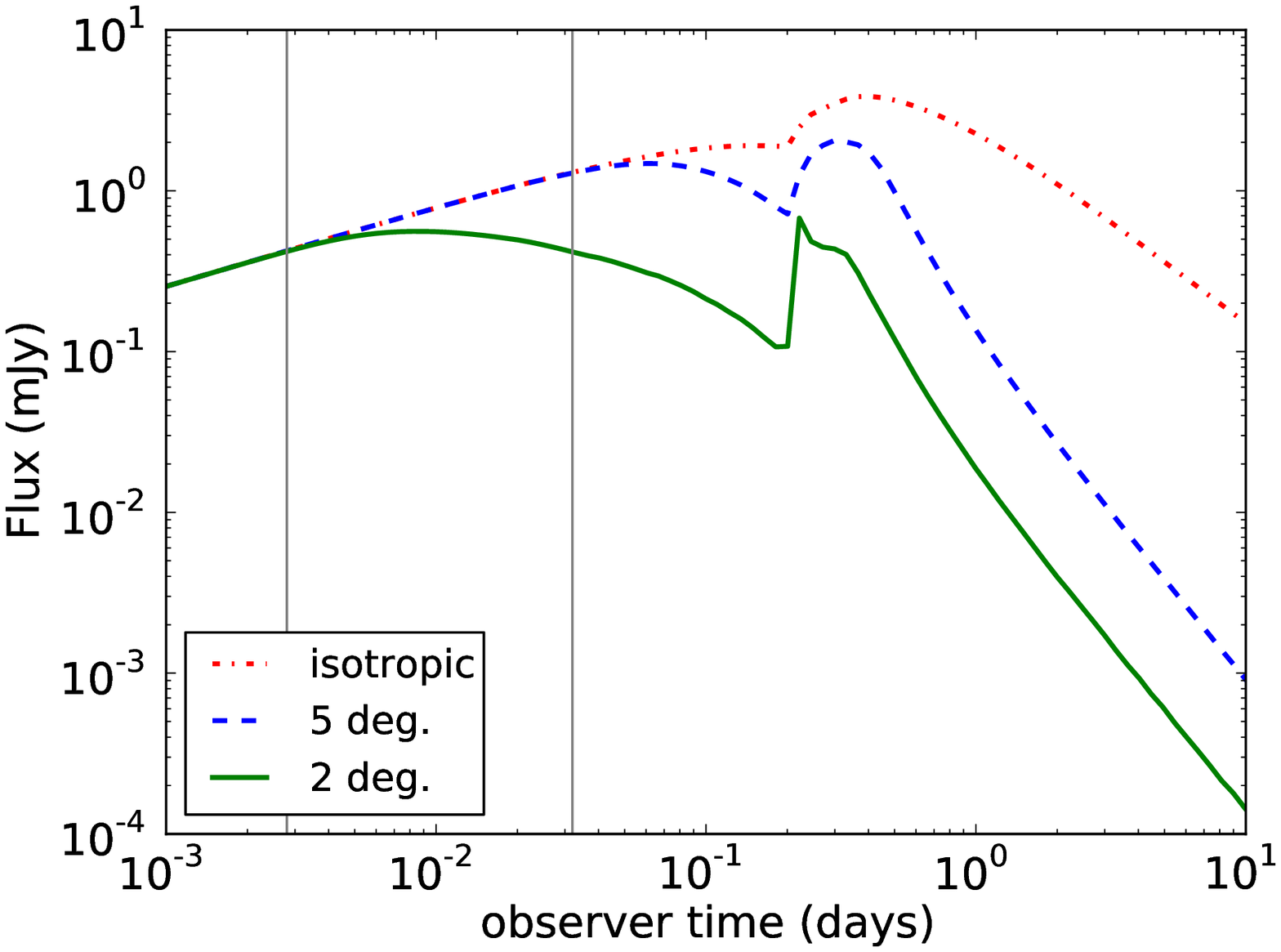}
 \includegraphics[scale=0.4]{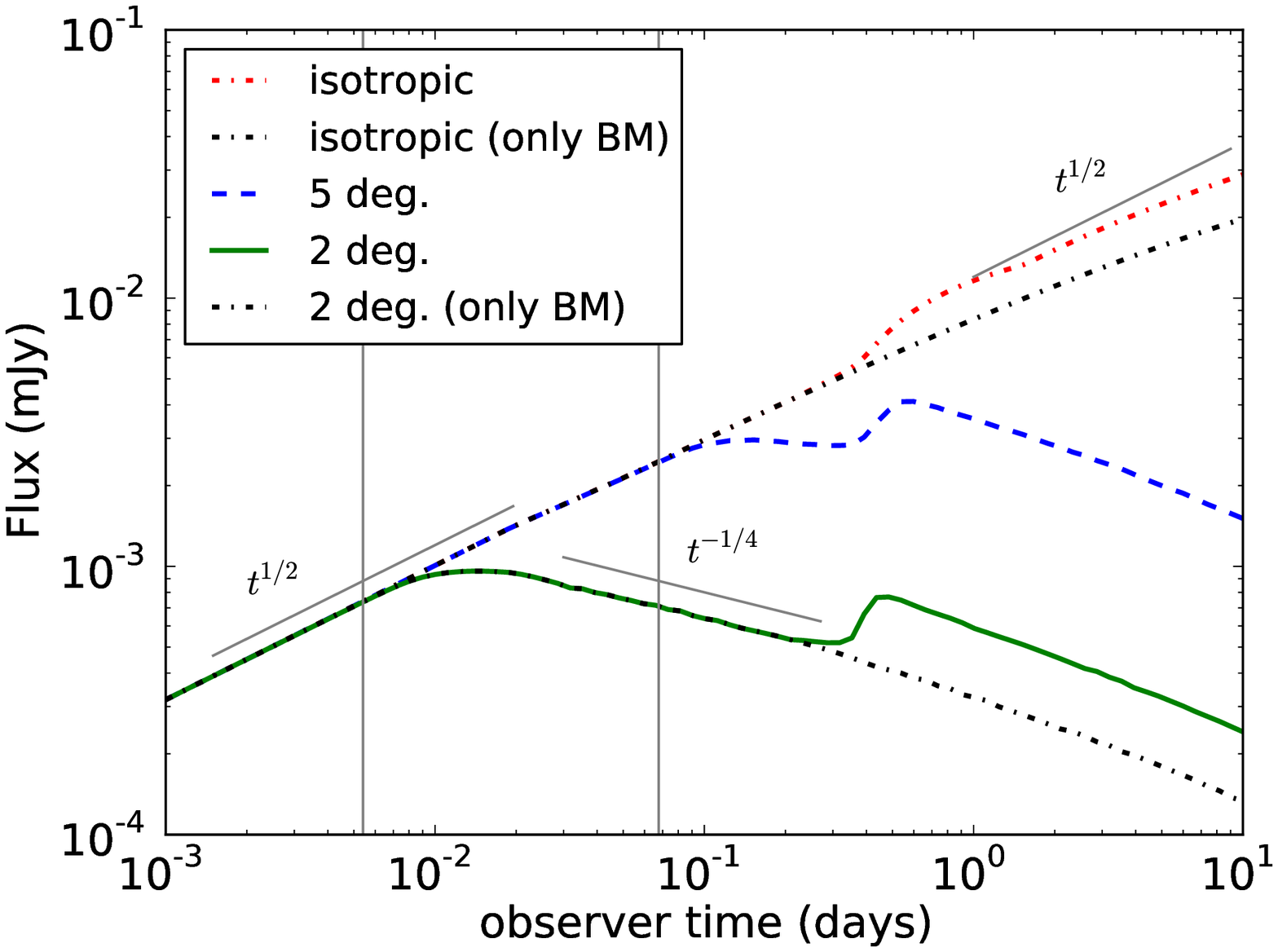} 
 \includegraphics[scale=0.4]{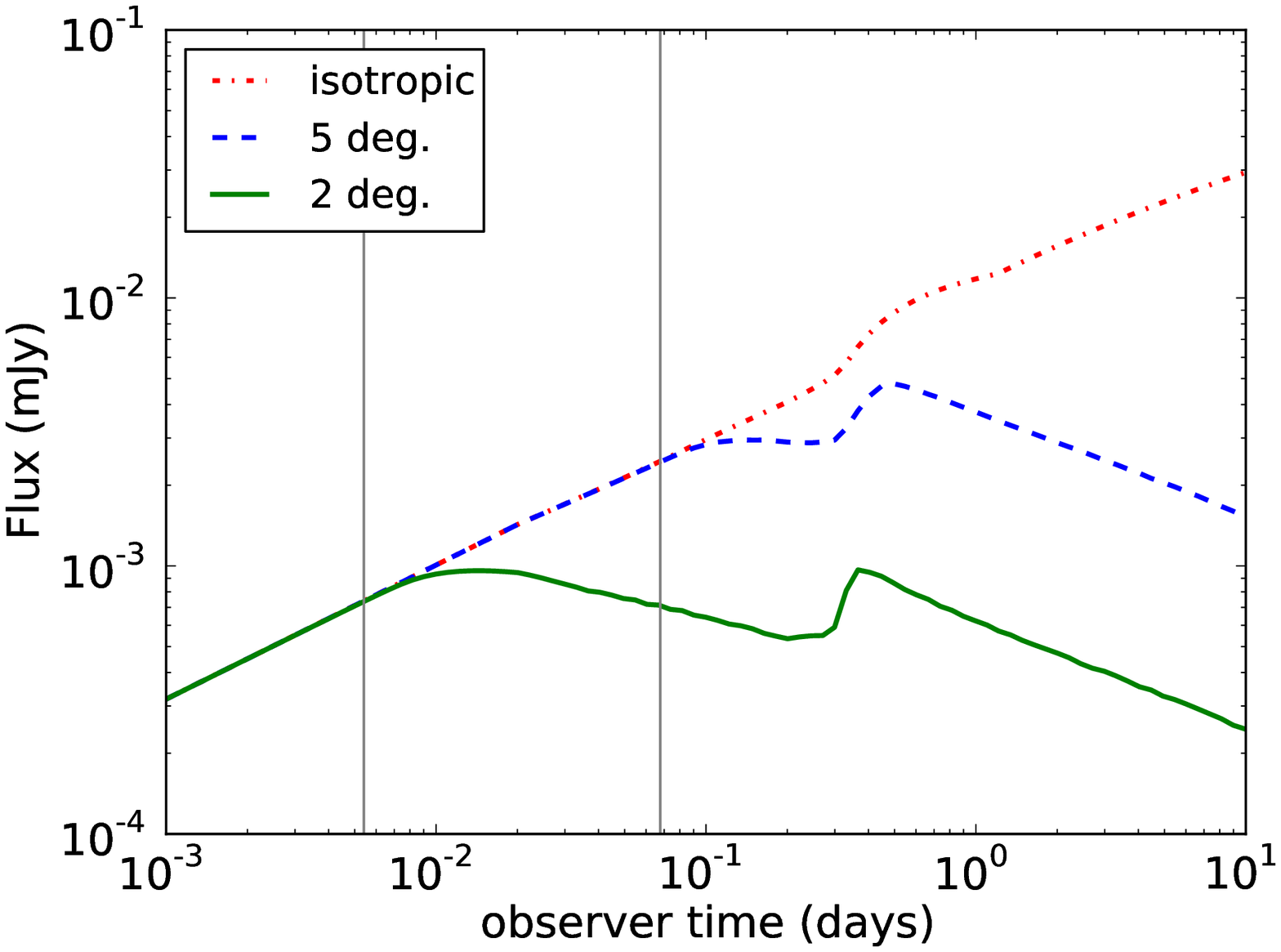}
\caption{Optical (top) and radio (bottom) light curves for case 1 (left) and case 2 (right) for different values of the jet opening angle $2\theta_{h}$. In all cases an increase of the flux due to energy injection from the second shell is observed. For small opening angles a flare appears in all four simulated cases which differs in shape according to the frequency. The peak flux is greater the higher the Lorentz factor and energy of the second shell are. The jet break estimation is denoted with a vertical line for both 2 and 5 degree jets. For case 1 we overplot the light curves produced for a single BM shell for a spherical explosion and a 2 degrees hard-edged jet. The analytical estimation of the slope is shown above the optical and radio light curves.}
\label{c910lc}
\end{figure*}

\begin{figure*}
 \centering
 \includegraphics[scale=0.4]{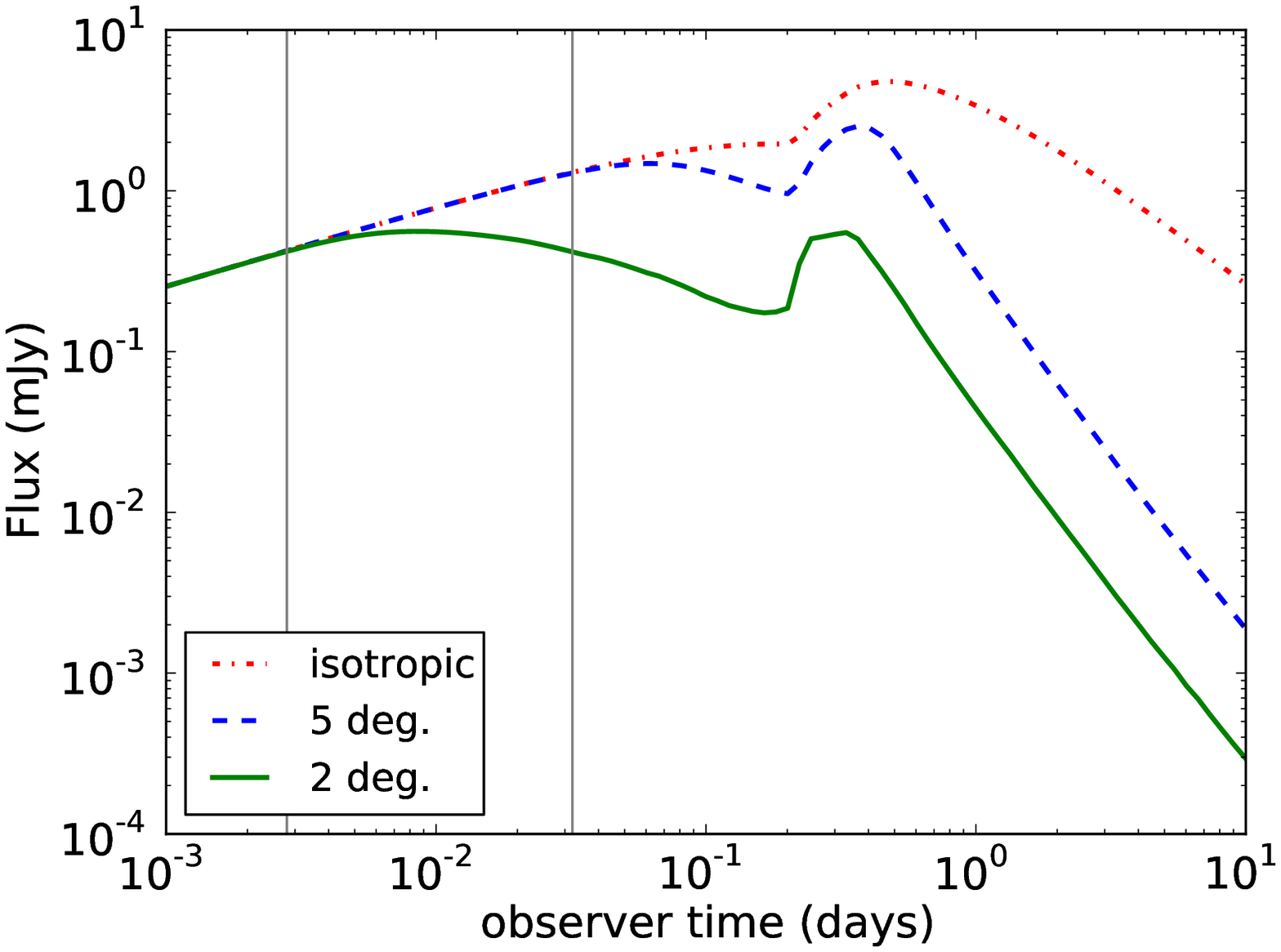}
 \includegraphics[scale=0.4]{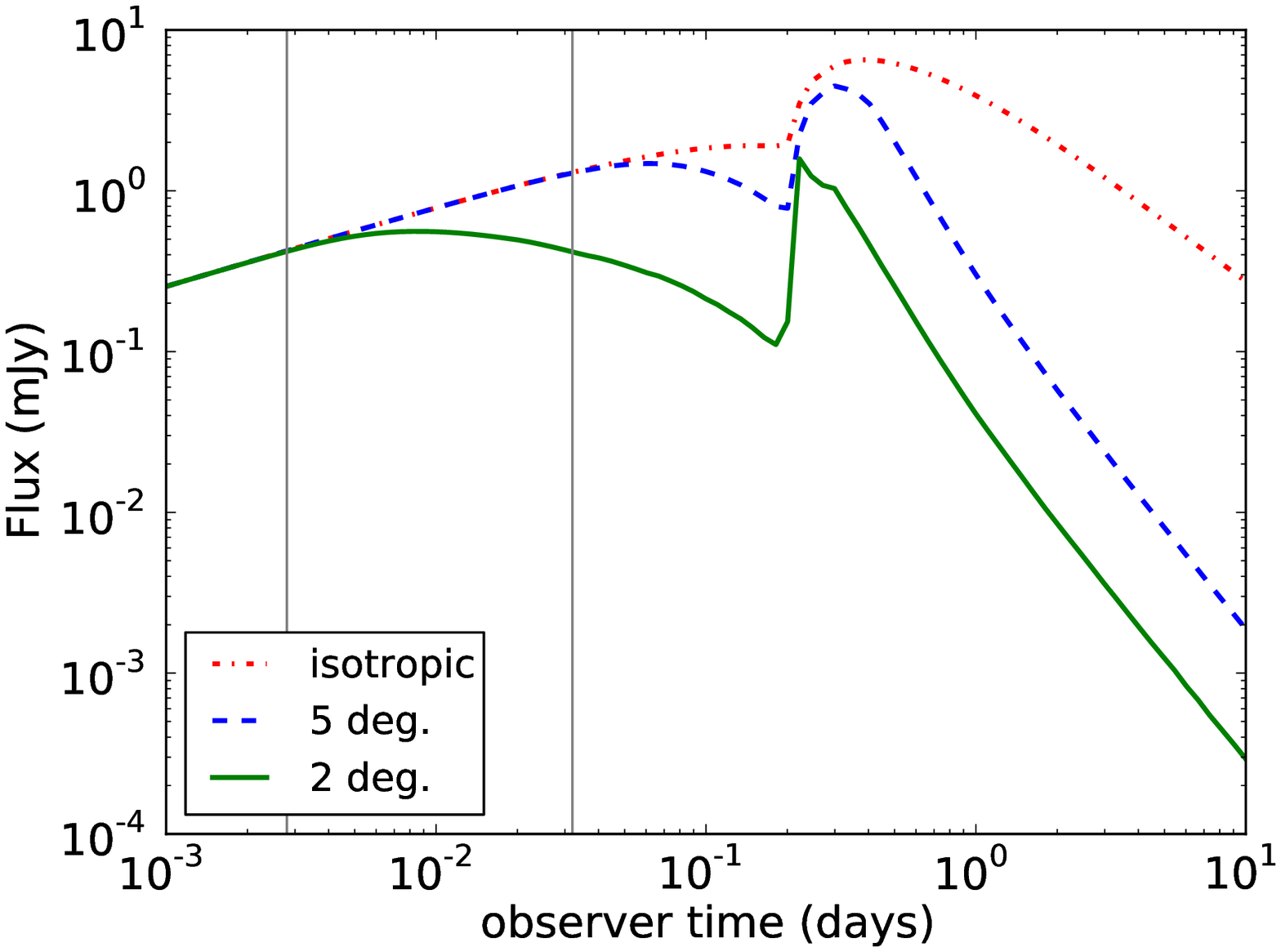}
 \includegraphics[scale=0.4]{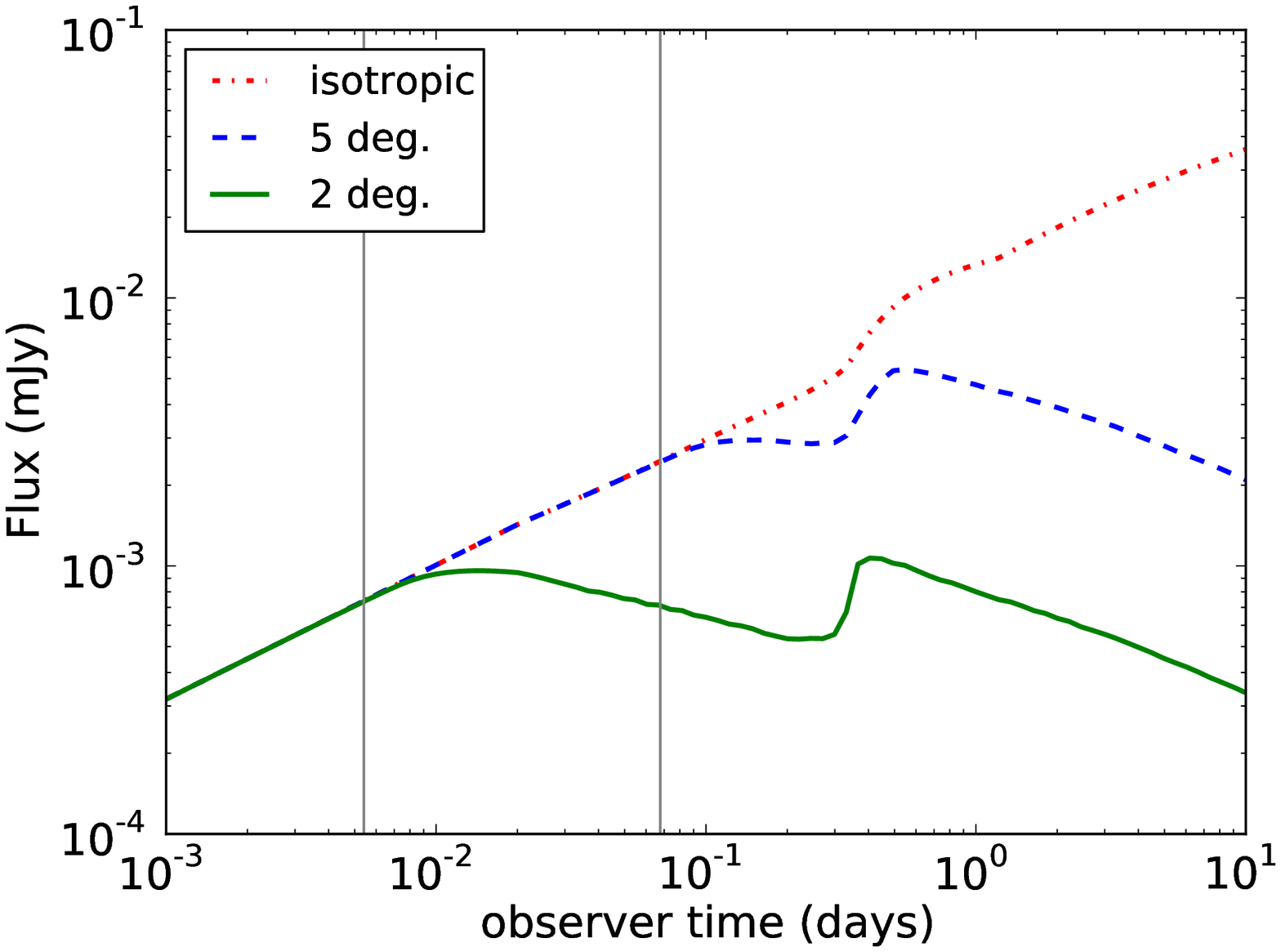}
 \includegraphics[scale=0.4]{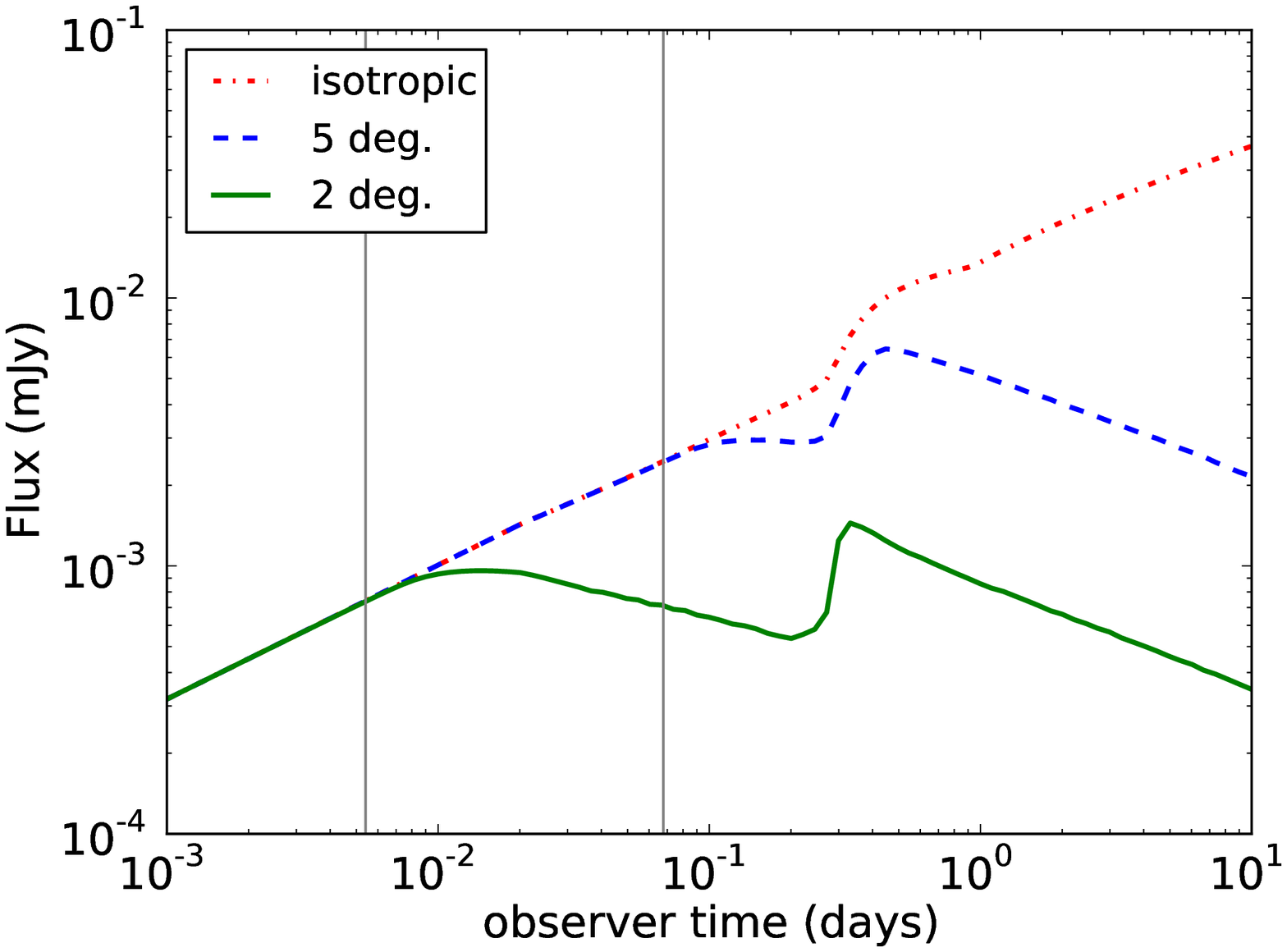}
\caption{Optical (top) and radio (bottom) light curves for case 3 (left) and case 4 (right) for different values of the jet opening angle $2\theta_{h}$. In all cases an increase of the flux due to energy injection from the second shell is observed. For small opening angles a flare appears in all four simulated cases which differs in shape according to the frequency. The peak flux is greater the higher the Lorentz factor and energy of the second shell are. The jet break estimation is denoted with a vertical line for both 2 and 5 degree jets.}
\label{c1112lc}
\end{figure*}

\subsection{Optical and radio light curves}

For the four cases described in section 2.3 we produce optical and radio light curves. We examine each case assuming different opening angles of the jet. In Figure \ref{c910lc} and \ref{c1112lc} we present the light curves produced by the radiation code for spherical and collimated expansion of the two colliding shells.

A rebrightening of the optical light curve is observed for $t_{obs}=0.2$ days for the spherical explosion in case 1 (Fig. \ref{c910lc}), which is attributed to the interaction of the two shells when the second one reaches the back of the BM shell. We find that the smaller the opening angle of the jet, the steeper the rise of the flare. This can be understood from the fact that for a jet with small opening angle there is, at a given observer time, less emission still arriving from earlier emission times and higher emission angles. As we show in section 3.5 (Figures \ref{eiflarec10}, \ref{eiflarec12}) early emission time contribution comes from high emission angles on the jet, thus the features of the light curve are less smeared out for a collimated outflow.

This can also explain the difference we observe in the decreasing rate of the flux at late observer times. Since there is no contribution from early emission times, the flux for a jetted outflow will decline faster compared to a spherical explosion. We also observe that the flare is sharper in case 2 compared to case 1 and in case 4 compared to all four cases (Fig. \ref{c910lc} and \ref{c1112lc}). This confirms our hypothesis that the flare is strongly dependent on the dynamical properties of the collision. For case 2, where the forward and reverse shock of the second shell are significantly stronger compared to case 1 (Fig. \ref{etrans}), the change in the flux during the collision of the two shells appears substantially stronger. In cases 3 and 4 we observe the same behaviour in the light curves. In case 4, the higher Lorentz factor together with the higher energy imposed in the second shell, lead to a flare clearly stronger compared to all the other cases for a 2 degrees opening angle of the jet. In that case the timescale of the variation for the flare is $\Delta T/T = 1.08$ for the optical and $\Delta T/T = 1.90$ for the radio, where $\Delta T$ is calculated as the full width at half maximum of the flare and \textit{T} is the observer time of the peak. The relative flux increase with respect to the underlying afterglow is $\Delta F/F = 3.95$ for the optical and $\Delta F/F = 1.04$ for the radio which is significantly reduced due to ssa mechanism. Although extensive comparison to observational data is beyond the scope of this work, we note that this case resembles the afterglow of GRB 060206 which shows an increase in brightness by $\sim 1$ mag 1 hr after the burst, followed by a typical broken power-law decay \citep{Stanek2006,Wozniak2006b}.

Before and after the flares the resulting light curves follow the analytically predicted slopes for a single forward shock as shown in Fig \ref{c910lc}. The light curves at a given frequency depend on the temporal evolution of the characteristic frequencies of the system, that is in our case the synchrotron peak frequency $\nu_{m}$ and the self-absorption frequency $\nu_{sa}$. As described in \citet{Granot2002} these slopes read $t^{1/2}$ for times before the break frequency $\nu_{m}$, crosses the observed frequency $\nu$ and $t^{3(1-p)/4}$ for the time after.  
In addition to this behaviour the slope of the light curve steepens by $t^{-3/4}$ after the jet break, in both optical and radio frequency, for small opening angles of the jet. This steepening appears clearer in the radio ($t_{obs,br}=0.0054$ days for 2 deg. jet and $t_{obs,br}=0.068$ days for 5 deg. jet) where the emission comes from late emission times and lower emission angles close to the jet axis. In the optical, where earlier emission times also contribute to the observed flux (as discussed in section 3.5 and shown in Fig. \ref{eiflarec10} and \ref{eiflarec12}), this steepening is delayed and appears at $t_{obs}=0.075$ days almost at the same time as the break frequency $\nu_{m}$ crosses the optical band.

Throughout our calculations and for all cases the critical frequencies satisfy $\nu_{c}>\nu_{obs}$, $\nu_{c}>\nu_{m}$ and $\nu_{c}\gg\nu_{sa}$ at the BM shock and the forward shock of the second shell, where $\nu_{c}$ is the cooling frequency (all critical frequencies are calculated using the formulas found in Table 2 of \citet{Granot2002} for a BM solution. Between $t_{obs}$ of 0.001 and 10 days, $\nu_{c}$ decreases from $10^{20}$ Hz to $10^{17}$ Hz, while $\nu_{m}$ decreases from $10^{16}$ Hz to $10^{10}$ Hz for the BM shock of case 1). Hence, for the frequencies discussed in the present work we are always in the slow cooling regime. In addition to the fact that the shock regions forming at the second shell are a lot thinner than a typical BM profile, this allows us to neglect, with small error, the effect of synchrotron cooling on the electron distribution for the times under consideration and consequently any changes on the self-absorption coefficient. 

Furthermore our assumption that inverse Compton (IC) doesn't influence the observed spectrum is confirmed as follows. There are two ways in which IC scattering can change the overall synchrotron spectral component. First by producing an additional emission component at high frequencies and second by dominating the electron cooling and thus reducing the available energy for synchrotron radiation. As discussed in \citet{Sari2001}, the first is estimated by considering the ratio of specific fluxes measured at the peak of the respective flux components, $f_{max}$ for the synchrotron and $f_{max}^{IC}$ for the IC respectively. Then this ratio is $f_{max}^{IC}/f_{max} \sim 1/3 \sigma_{T}nR$, where $\sigma_{T}$ is the Thompson cross-section and $n$ the electron density at distance $R$. This ratio remains well below unity for all the shock surfaces in all four simulations which are taken into consideration. The second factor can be estimated directly from the ratio $\eta\epsilon_{E}/\epsilon_{B}$. As shown by Sari \& Esin, the IC cooling rate will be unimportant compared to synchrotron if $\eta\epsilon_{E}/\epsilon_{B}\ll1$, where $\eta=(\gamma_{c}/\gamma_{m})^{2-p}=(\nu_{c}/\nu_{m})^{-(p-2)/2}$. In our case this ratio is close to unity for the early afterglow but decreases rapidly and remains below unity during our simulation covering the pre and post flaring activity period.

\begin{figure}
 \centering
 \includegraphics[scale=0.43]{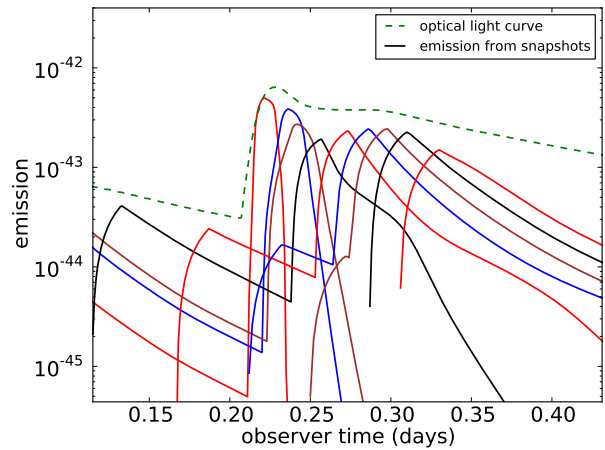}
\caption{Optical light curve and single emission images for several snapshots of the two shell system for case 4. The emission time of the snapshots range from $t_{e}=7.64\times10^{6}$ s for snapshot 50 to $t_{e}=1.26\times10^{7}$ s for snapshot 350.}
 \label{single}
\end{figure}
\subsection{The shape of the optical flare.}

The flaring activity as observed in the optical follows three stages. At the first stage the Lorentz factor and the thermal energy of the BM matter in front of the second shell increase due to the propagation of the forward shock leading to the sudden rise of the flux. The flare observed in Fig. \ref{c910lc} and \ref{c1112lc}, is attributed to that increase. 
This behaviour continues until $t_{obs}=0.23$ days when the reverse shock crosses the back of the second shell. At that time the forward shock starts decelerating and that can be seen as a reduction of the flux, after the initial peak, in the optical light curves. At time $t_{obs}=0.25$ days the shocked BM matter has separated from the shell and now propagates in the BM shell while constantly heating the swept up matter. The flux for that stage of the motion remains almost constant in the optical light curve as the density in front of the forward shock continuously increases and the heated matter compensates in the flux for the deceleration of the forward shock. This remains until the forward shock of the second shell and the BM shell merge at time $t_{obs}=0.31$ days. At that time the optical light curve presents a change in the slope and the flux starts decreasing as the emission now originates from the merged shell while propagating into the ISM. 

In figure \ref{single} we plot the optical light curve for case 4 considering a hard-edged jet with opening angle $2\theta_{h}=2$ during the period of the flare and the emission from several snapshots before, after and during the collision of the two shells. The emission from each snapshot consists of a BM shell contribution which peaks at early observer times and the contribution from the second shell which peaks during the flare and at later times. A comparison between the light curve and the emission snapshots shows that immediately after the forward shock of the second shell is created the flare is observed ($t_{e}=7.64\times10^{6}$ s). The stage of the forward shock deceleration and the flux reduction can be seen from $t_{e}=7.64\times10^{6}$ to $t_{e}=1.01\times10^{7}$ s. As soon as the forward shock separates from the second shell, the shape of the emission arising from the forward shock becomes sharper ($t_{e}=1.05\times10^{7}$ s). The third stage of the flare where the forward shock propagates inside the BM matter shapes the plateau which is observed from $t_{obs}=0.25$ to $t_{obs}=0.31$ days and can be seen from $t_{e}=1.05\times10^{7}$ to $t_{e}=1.26\times10^{7}$ s. At that emission time the merger of the two shells has almost completed and the flux starts decaying.

\subsection{Time delay observed between optical and radio flares.}

Comparing the optical to the radio light curves, we notice that the flaring activity occurs with a distinct time delay (approximately 0.1 days) for the latter ones (Fig. \ref{optradiocompare}). 
The reason for this is the ssa mechanism. For optical emission the jet is optically thin and the contribution from the second shell is obtained as soon as the forward shock is created while propagating in the BM shell. Below the self-absorption frequency though the jet behaves differently. In the radio the jet is optically thick due to ssa mechanism and the merger becomes visible only after the collision has nearly completed. This results in observing the plateau at the second stage of the optical flare and a sharp peak in the radio flare. 
This highlights the significance of taking into account ssa when calculating radio light curves. 
In fact any variability resulting from changes in the fluid conditions may manifest in a chromatic fashion \citep{HvE10b}. For that reason the jet break is significantly postponed in the radio light curves (Figs. \ref{c910lc} and \ref{c1112lc}).

\subsection{Emission images}

\begin{figure}
\centering
\includegraphics[scale=0.4, angle=270]{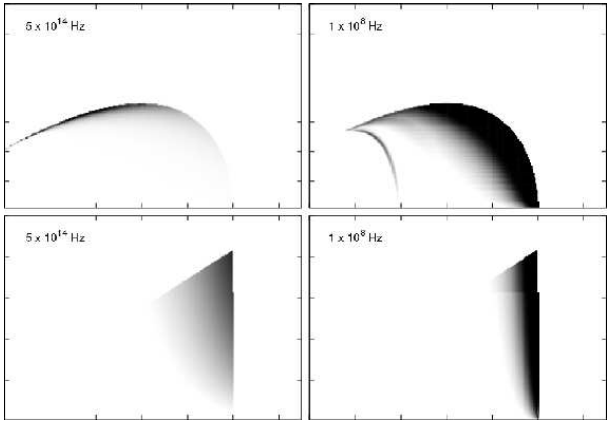}
\caption{Optical $5\times10^{14}$ Hz, (left) and radio $10^{8}$ Hz (right) emission image in case 4 at time $t_{obs}=10$ days. The upper pictures correspond to a spherical explosion and the lower ones to a jet with opening angle $2\theta_{h}=2$. The horizontal axis scales from $2\times10^{17}$ to $1.5\times10^{18}$ cm and the vertical from $5\times10^{15}$ to $3.5\times10^{16}$ cm. The main contribution area for the optical image comes from higher emission angles while for the radio image lower emission angles contribute the most. For a hard-edged jet (lower graphs) the radio image appears stronger, since the main contribution area to the optical at the back of the jet is not taken into account. The remainder of the second shell significantly heated after the traverse of the reverse shock reveals its contribution to the radio image.}
\label{eic12}
\end{figure}

\begin{figure}
\centering
\includegraphics[scale=0.42, angle=270]{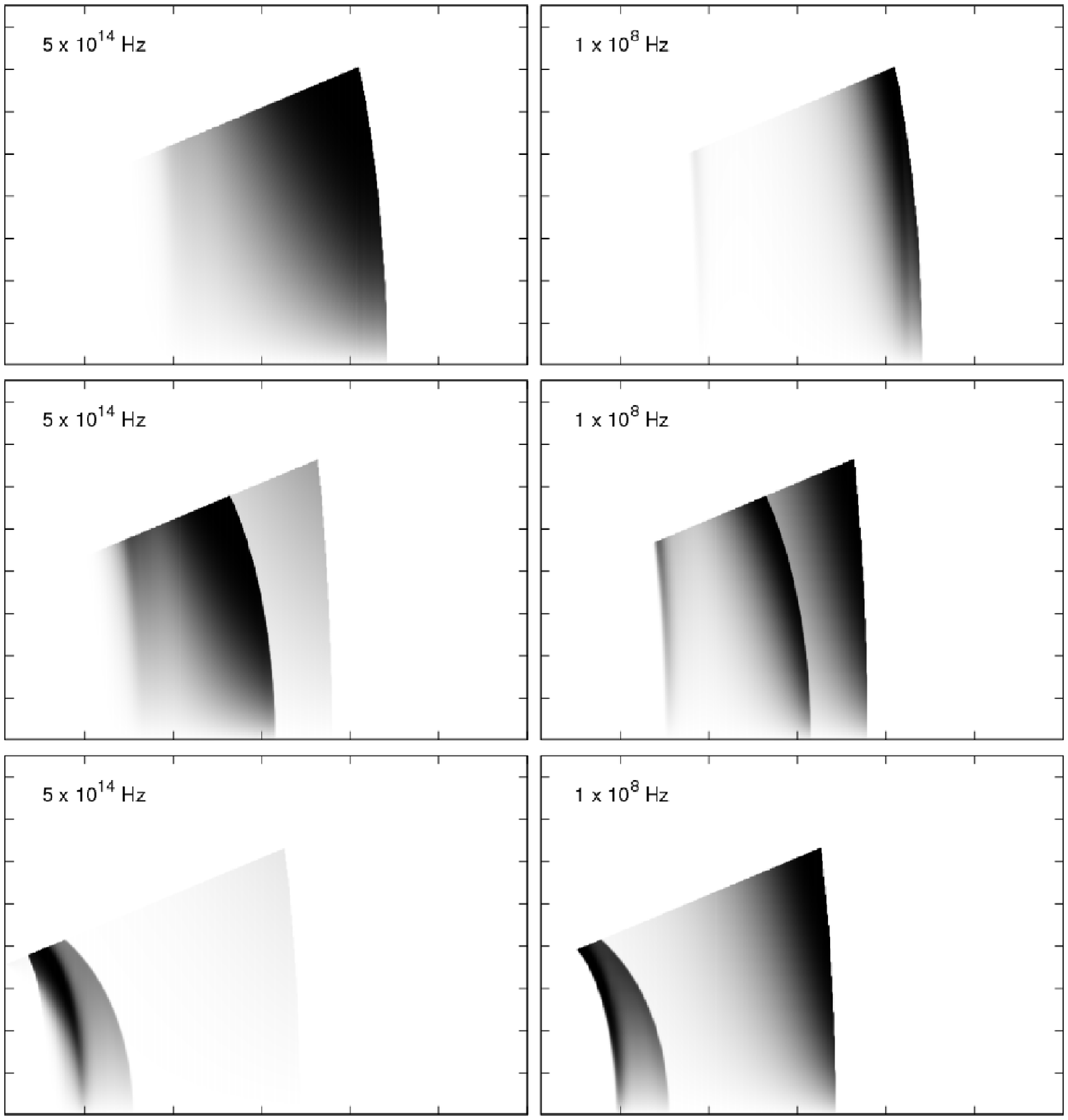}\vspace{5pt}
\includegraphics[scale=0.42, angle=270]{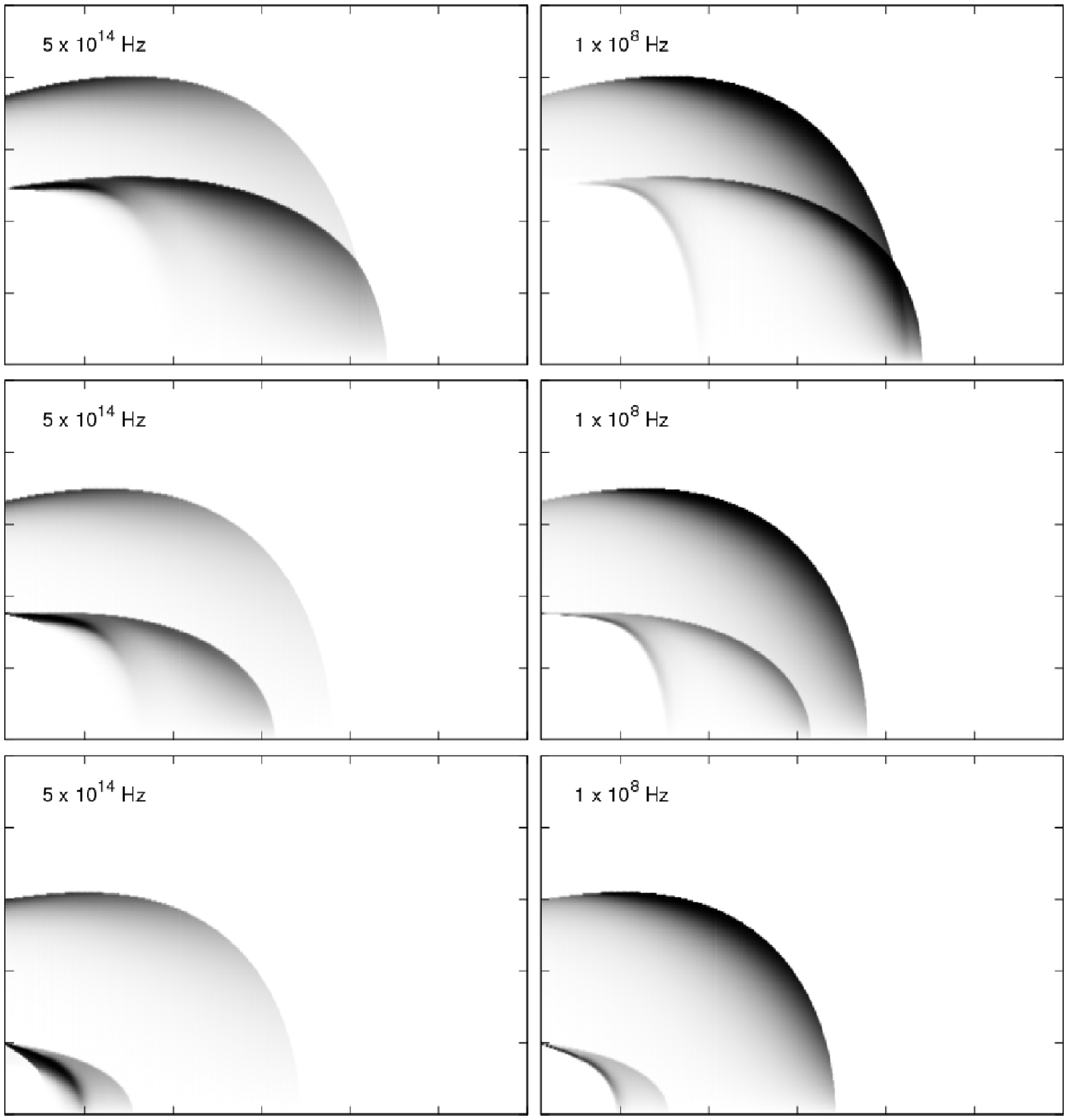}\vspace{5pt}
\caption{Optical ($5\times10^{14}$ Hz) and radio ($10^{8}$ Hz) emission images of the two-shell system during the flaring activity for case 2 and jet opening angle 2 degrees (top figures) and 20 degrees (bottom figures). The horizontal axis scales from $2\times10^{17}$ to $5\times10^{17}$ cm and the vertical from $2\times10^{13}$ to $8.5\times10^{15}$ cm for the 2 degrees jet and from $2\times10^{13}$ to $2.5\times10^{16}$ cm for the 20 degrees jet. For each subfigure the bottom images correspond to $t_{obs}=0.23$ days, which is the time the sudden rise of the flux is observed, the middle ones to the weak decay at $t_{obs}=0.28$ days corresponding to the propagation of the forward shock into the BM medium, and the top ones to the fast decay at $t_{obs}=0.35$ days, once the merger has completed.}
\label{eiflarec10}
\end{figure}

\begin{figure}
\centering
\includegraphics[scale=0.42, angle=270]{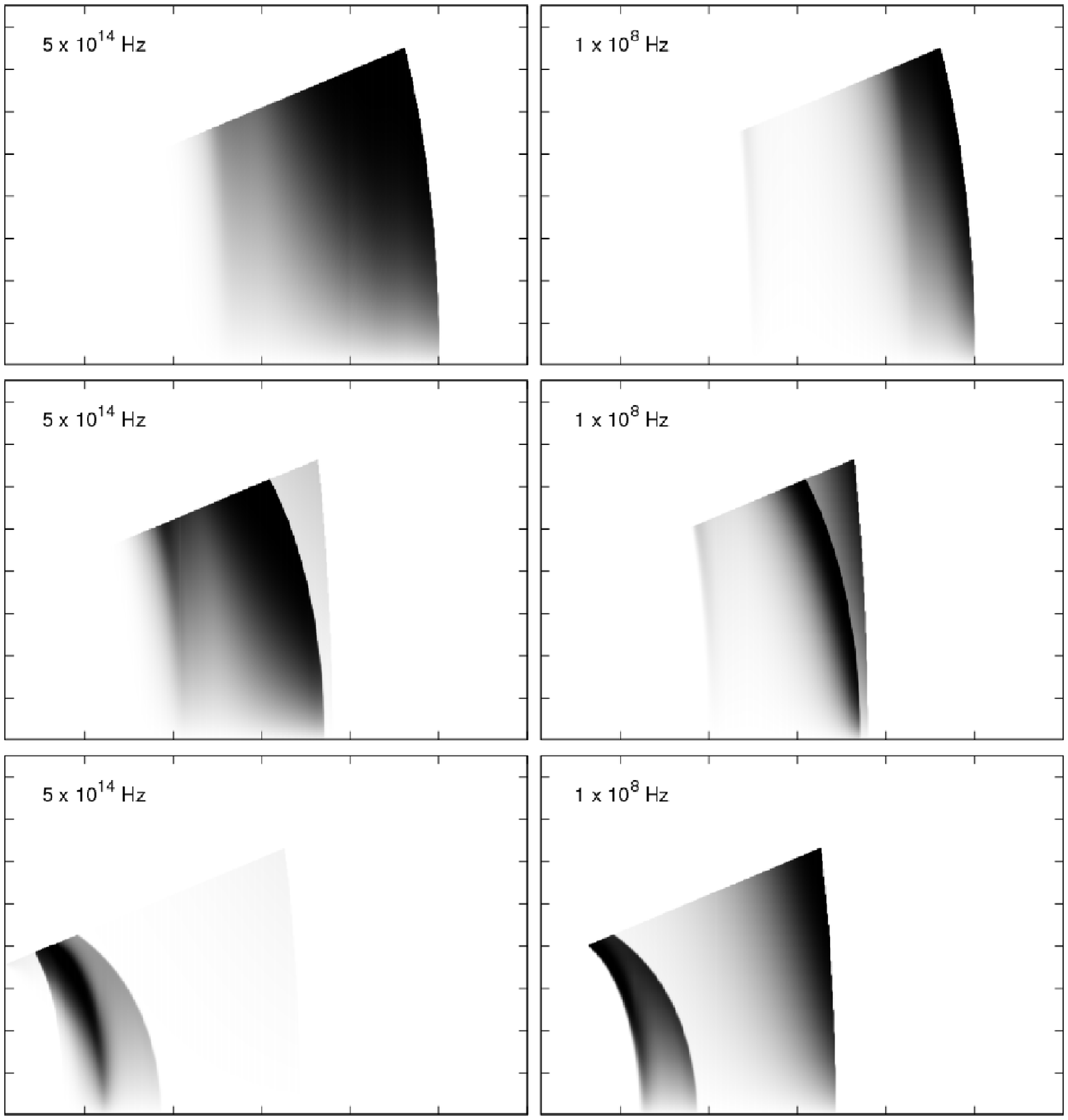}\vspace{5pt}
\includegraphics[scale=0.42, angle=270]{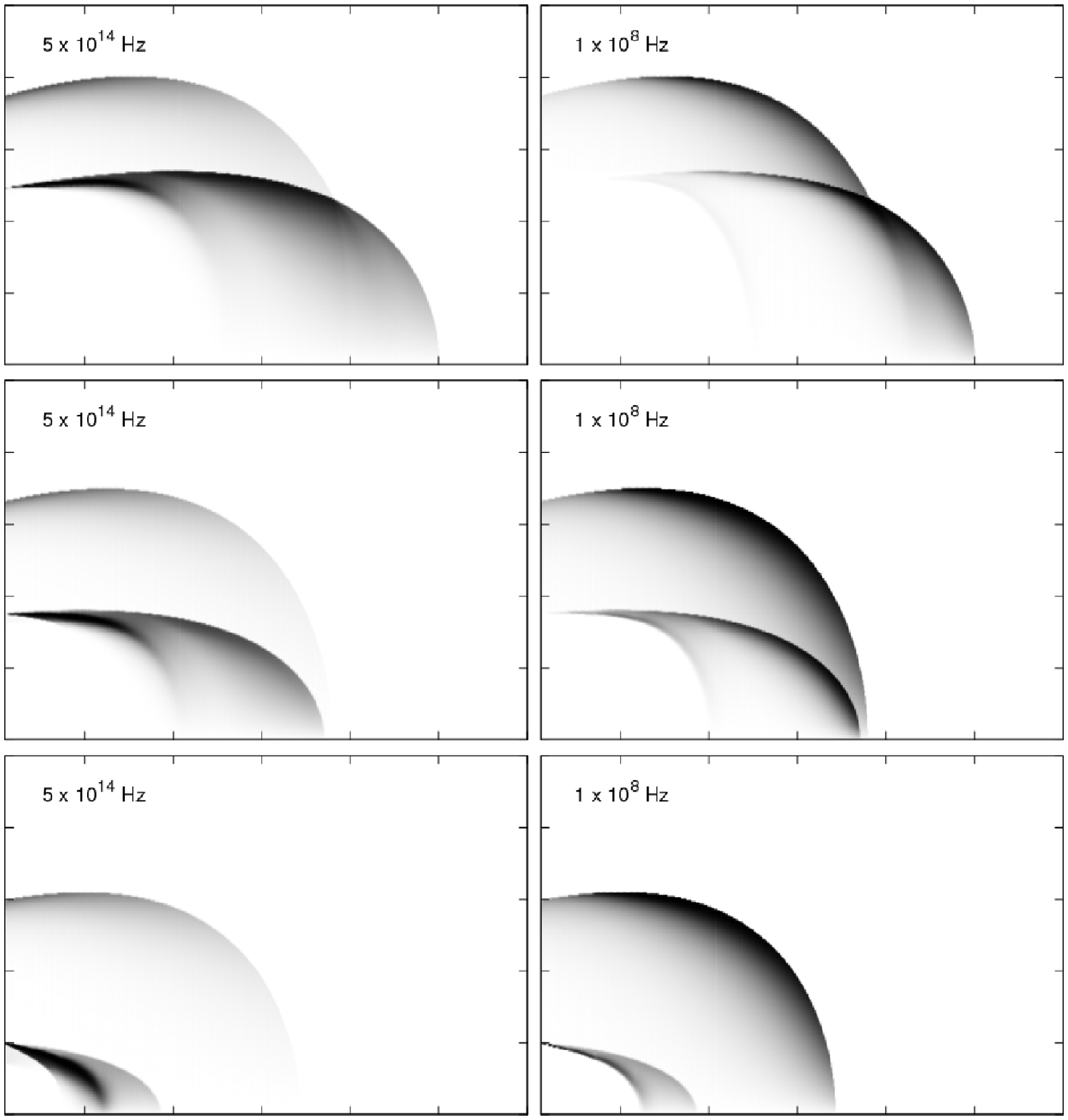}\vspace{5pt}
\caption{Optical ($5\times10^{14}$ Hz) and radio ($10^{8}$ Hz) emission images of the two-shell system during the flaring activity for case 4 and jet opening angle 2 degrees (top figures) and 20 degrees (bottom figures). The horizontal axis scales from $2\times10^{17}$ to $5\times10^{17}$ cm and the vertical from $2\times10^{13}$ to $8.5\times10^{15}$ cm for the 2 degrees jet and from $2\times10^{13}$ to $2.5\times10^{16}$ cm for the 20 degrees jet. For each subfigure the bottom images correspond to $t_{obs}=0.23$ days, which is the time the sudden rise of the flux is observed, the middle ones to the weak decay at $t_{obs}=0.28$ days corresponding to the propagation of the forward shock into the BM medium, and the top ones to the fast decay at $t_{obs}=0.35$ days, once the merger has completed.}
\label{eiflarec12}
\end{figure}

By directly plotting the relative contributions to the light curve from different parts of the fluid, the reasons for the differences between radio and optical light curves become even more obvious. 
The flux at a given observer time is obtained by solving the linear radiative transfer equation through the evolving fluid for a large set of rays. By separately storing the local contributions to the emerging rays, we have created emission images showing exactly the relative contributions of different parts of the jet.
We produce emission images for different observer time and opening angle, for optical ($5\times10^{14}$ Hz) and radio ($10^{8}$ Hz) frequencies covering the time before, during and after the flaring activity is observed. In all the images the jet axis is aligned to the horizontal direction and the observer is at the right end of the horizontal axis.
Which area of the jet is the main contributing area to the emission is strongly frequency dependent. When the frequency of observation lies well above the self absorption frequency, the system is optically thin and the main contribution to the emission is from early emission times and from high emission angles on the jet.  
For lower frequencies the system becomes optically thick due to self absorption, hence the main contribution to the emission is from later emission times and the emitting region shifts to lower emission angles closer to the jet axis. 
This behaviour is observed throughout the figures \ref{eic12}-\ref{eiflarec12}, where we plot the ring-integrated, absorption-corrected local emission coefficients, as a difference in the contrast between optical and radio emission.

\begin{figure}
 \centering
\includegraphics[scale=0.43]{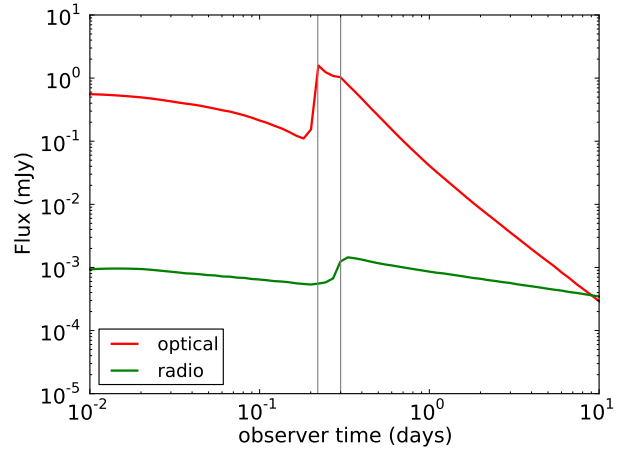}
\caption{Optical and radio light curve comparison for case 4, assuming a jet opening angle $2\theta_{h}=2$. The optical flare precedes the   radio one by 0.1 days. The three different stages of the flare evolution can be seen on the optical flare.}
 \label{optradiocompare}
\end{figure}

In Fig. \ref{eic12} we plot the optical and radio emission images at observer time $t_{obs}=10$ days, long after the merger has completed, for case 4 for a spherical explosion and for a hard-edged jet of opening angle $2\theta_{h}=2$. We observe that different parts of the merged shell contribute to different parts of the spectrum. The optical contribution emerges mainly from high emission angles of the merged shell whereas radio emission comes mainly from lower emission angles close to the jet axis. This behaviour is also encountered in the emission images derived for the hard-edged jet. In this case, high emission angles are excluded from the calculation of the emission, which affects the optical more strongly than the radio emission. 
A contribution from the remainder of the second shell is observed at the radio emission image of the spherical explosion. Although significantly reduced due to ssa and almost two orders of magnitude less than the emission originating from the merged shell, this contribution owns its existence to the traverse of the reverse shock (Fig. \ref{eic12}). As the reverse shock crosses the second shell the Lorentz factor decreases, and by time $t_{obs}=10$ days, when the shell has been significantly decelerated, the peak frequency of the synchrotron spectrum originating from that region is shifted to lower frequencies.
The stronger the reverse shock the stronger the deceleration of the shell. The contribution of the second shell is therefore expected to be higher.

In Fig. \ref{eiflarec10} we show the evolution of the two shells before, during and after the collision phase for case 2, for both radio and optical frequencies assuming opening angles of the jet 2 and 20 degrees (rather than 5, to better capture high-angle emission and illustrate the contrast between narrow and wide jets). The collision appears between observer times $t_{obs}=0.28$ and $t_{obs}=0.35$ days which verifies that the break in the light curves occurs at the same time as the collision of the two shells. 
From the emission images it becomes clear that the smearing out of the flare from early emission time contribution becomes less the smaller the opening angle of the jet is. Comparing the two different cases in figure \ref{eiflarec10} (case 2) and figure \ref{eiflarec12} (case 4) it is evident that in the optical case and for large opening angles, early time contribution dominates the emission and drowns out the flaring activity. For small opening angles this effect is suppressed revealing this way the contribution of the second shell while it collides with the BM shell. 
In the radio emission images, where the main contributing area to the emission is transferred to lower emission angles, we do not observe this behaviour. Instead as described in section 3.2, in the radio frequency the jet is optically thick due to ssa mechanism and the merger can be seen only when the collision has almost completed. This difference in behaviour between optical and radio, manifests itself as a sharp rise in the radio light curve and a plateau at the flare in the optical light curve (see also Fig. \ref{optradiocompare}). 

From the emission images in fig. \ref{eiflarec10} and for a jet with an opening angle 20 degrees we derive the conclusion that for higher frequencies (optical, X-rays), for which the main contributing region to the observed flux shifts to higher emission angles on the jet, a double ring shaped image should appear in GRB late afterglow observations. In this type of image the inner ring carries information from the forward shock emission of the second shell, whereas the outer ring is determined from the emission at earlier stages of the forward shell. For radio frequencies where the observed flux arrives mainly from emission angles close to the jet axis, the distinct contribution of the two shells can only be observed after the merger has completed.

\section{Discussion and Conclusions}

We performed high resolution numerical simulations of late collisions between two ultra relativistic shells and produced optical and radio light curves and emission images for a spherical explosion case and different opening angles of a hard-edged jet. The AMR technique allowed us to reach high resolution and properly resolve the shocks developed during the merger of the shells.

The simulations have shown that different values of the Lorentz factor and energy of the second shell can significantly change the characteristics of the variability on the light curves. We demonstrate that for small opening angles, for which the flux is not smeared out by early emission time contribution, a flare appears at the light curve. The onset of the flare is found to be strongly dependent on the strength of the forward and reverse shock of the second shell and to become stronger the higher the Lorentz factor and energy of the second shell are. For case 4 in which the Lorentz factor and energy of the second shell are the highest in our simulations, the relative increase of the flux in respect to the underlying afterglow is $\Delta F/F = 3.95$ for the optical and $\Delta F/F = 1.04$ for the radio. The timescale of the variation of the flare is $\Delta T/T = 1.08$ and $\Delta T/T = 1.90$ respectively. The shape of the flare is understood through the dynamical simulations and the different stages at the light curve are associated with different parts of the collision process. We show that the difference in shape between the optical and radio flare as well as the time difference observed between them is a direct result of the ssa mechanism and the angle dependence of the emission. 
We predict that this type of behaviour should appear in late afterglow observations as a two-ring feature in spatially resolved optical emission images, although for the time being the resolution required would be unrealistic.
Athough a detailed numerical approach is required to fit observational data with the numerical results, a straightforward comparison was made between case 4 of our simulations and the optical flare observed in the afterglow of GRB 060206 showing strong resemblance in the magnitude and time variation of the flare. New low-frequency facilities such as ALMA are expected to provide enough data to compare with the radio light curves as well.
The details which determine the jet collimation as the jet propagates into the circumburst medium are currently not well understood. In this work we have used the same opening angle for both shells and we defer analysis of a scenario with different opening angles to future work. Such a study would require detailed simulations in 2D.
%A detailed 2D study is required to investigate both the effects of different opening angles of the two ejecta, as well as the spreading that may occur when the opening angle of the jet is extremely small, as in the case of $\theta_{jet}=2$ degrees considered later in this paper.

We have not explicitly discussed X-ray flares, although for these the largest amount of observational data is available. The X-ray emission is influenced by electron cooling (i.e. lies above the cooling break in the synchrotron spectrum) and is therefore also more sensitive to the details of particle acceleration than optical and radio emission. Especially in the case of multiple shocks, it becomes difficult to implement an approach to radiation that correctly captures all relevant physics in a simple parametrization. 
%Nevertheless, there is no reason to assume that the results obtained for the optical flares will not apply to X-ray flares as well. 
Many afterglows show X-ray flares superimposed on a broken power-law where a single slope describes the region before and after the flare and are often characterized by small timescale variations, $\Delta T/T \ll 1$ (e.g. Burrows et al. 2007). However, there are occasions, such as the GRB 060206, where the afterglow light curve presents a clear difference between the pre and post flare region with a significantly increased timescale variation, indicating an energy injection to the external shock as a probable cause for the generation of the flare \citep{Monf2006}. 
%For a detailed numerical study exploring X-ray flares in the afterglow we refer to future work. 
% X-ray flares appear in general in smaller timescales compared to the optical ones ($\Delta T/T \sim 0.1$) and in theory can be approached within the frameworks of the current model by applying a thinner and faster second shell maintaining unaltered the viability of the general concept that late time engine activity leads to a flare in the light curve. 
%The X-ray flares might be steeper due to the smaller sizes of the contributing emission regions of the merging shocks, but the viability of the general concept where late time engine activity leads to a flare in the light curve remains unaltered.

\section{Acknowledgments}
We thank R.A.M.J. Wijers and Petar Mimica for useful discussion. Computations performed on JADE (CINES) in DARI project c2010046216 and on HPC VIC3 at K.U.Leuven. These results were obtained in the framework of the Project GOA/2009/009 (K.U.Leuven). ZM acknowledges support from HPC Europa, project 228398. HJvE is supported by NASA under Grant No. 09-ATP09-0190 issued through the Astrophysics Theory Program (ATP).

\label{lastpage}

\end{document}